\documentclass[useAMS,usenatbib]{mn2e}

\usepackage{times}
\usepackage{amsmath,amssymb}
\usepackage{graphicx}
\usepackage{ctable}
\usepackage{comment}

\newcommand{\apj}{ApJ}
\newcommand{\apjs}{ApJS}
\newcommand{\aj}{AJ}
\newcommand{\aap}{A\&A}
\newcommand{\aaps}{A\&AS}
\newcommand{\mnras}{MNRAS}
\newcommand{\pasp}{PASP}
\newcommand{\icarus}{Icarus}
\newcommand{\nat}{Nature}
\newcommand{\apss}{Ap\&SS}
\newcommand{\araa}{ARA\&A}

\title[Search for transiting planets around M dwarfs: A pilot study]
{Photometric transit search for planets around cool stars from the western Italian Alps: A pilot study}
\author[Giacobbe et al.]{P. Giacobbe$^1$, M. Damasso$^{2,3}$, A. Sozzetti$^{4}$\thanks{E-mail:
sozzetti@oato.inaf.it}, G. Toso$^2$, M. Perdoncin$^5$, P. Calcidese$^2$, \newauthor 
A. Bernagozzi$^2$, E. Bertolini$^2$,  M. G. Lattanzi$^4$, and R. L. Smart$^4$\\
$^{1}$Dept. of Physics, University of Trieste, Via Tiepolo 11, I-34143 Trieste, Italy \\
$^{2}$Astronomical Observatory of the Autonomous Region of the Aosta Valley, Loc. Lignan 39, 11020 Nus (Aosta), Italy \\
$^{3}$Dept. of Astronomy, University of Padova, Vicolo dell'Osservatorio 3, I-35122 Padova, Italy \\
$^{4}$INAF - Osservatorio Astronomico di Torino, Via Osservatorio 20, I-10025 Pino Torinese, Italy \\
$^{5}$Dept. of Physics, University of Torino, Via Giuria 1, I-10125 Torino, Italy}
\begin{document}

\date{Accepted ????. Received ???}

\pagerange{\pageref{firstpage}--\pageref{lastpage}} \pubyear{2012}

\maketitle

\label{firstpage}

\begin{abstract}
We present the results of a year-long photometric monitoring campaign of a sample of 23 nearby ($d<60$ pc), 
bright ($J<12$) dM stars carried out at the Astronomical Observatory of
the Autonomous Region of the Aosta Valley, in the western Italian Alps. 
This program represents the `pilot study' for a long-term photometric transit search for 
planets around a large sample of nearby M dwarfs, due to start 
with an array of identical 40-cm class telescopes by the Spring of 2012. 
In this study, we set out to $a)$ demonstrate 
the sensitivity to $<4$ R$_\oplus$ transiting planets with periods of a few days around our 
program stars, through a two-fold approach that combines a characterization of the 
statistical noise properties of our photometry with the determination of transit detection 
probabilities via simulations, and $b)$ where possible, improve our knowledge 
of some astrophysical properties (e.g., activity, rotation) of our targets by combining 
spectroscopic information and our differential photometric measurements. We achieve a typical nightly RMS 
photometric precision of $\sim5$ mmag, with little or no dependence on the instrumentation used or on 
the details of the adopted methods for differential photometry. The presence of correlated (red) noise in our data 
degrades the precision by a factor $\sim1.3$ with respect to a pure white noise regime. 
Based on a detailed stellar variability analysis, 
$a)$ we detected no transit-like events (an expected result given the sample size); $b)$ we determined 
photometric rotation periods of $\sim$0.47 days and $\sim$0.22 days for LHS 3445 and GJ 1167A, respectively; 
$c)$ these values agree with the large projected rotational velocities ($\sim25$ km/s and 
$\sim33$ km/s, respectively) inferred for both stars based on the analysis of 
archival spectra; $d)$ the estimated inclinations of the stellar rotation axes for LHS 3445 and GJ 1167A 
are consistent with those derived using a simple spot model; $e)$ short-term, low-amplitude flaring events were recorded 
for LHS 3445 and LHS 2686. Finally, based on simulations of transit signals of given period and 
amplitude injected in the actual (nightly reduced) photometric data for our sample, we derive 
a relationship between transit detection probability and phase coverage. We find that, using the BLS 
search algorithm, even when phase coverage approaches 100\%, 
there is a limit to the detection probability of $\approx 90\%$. Around program stars with phase coverage $>50\%$
we would have had $>80\%$ chances of detecting planets with $P<1$ day inducing fractional transit depths $>0.5\%$, 
corresponding to minimum detectable radii in the range $\sim1.0-2.2$ $R_\oplus$. 
These findings are illustrative of our high readiness level ahead of the main survey start.

\end{abstract}


\section{Introduction}\label{intro}

M dwarf stars, with masses $M_\star\leq0.6$ M$_\odot$, make up the vast 
majority of the reservoir of nearby stars within $\sim 25-30$ pc. These stars have not traditionally 
been included in large numbers in the target lists of 
radial-velocity (RV) searches for planets for two main reasons: 1) their
intrinsic faintness, which prevented Doppler surveys in the optical from
achieving very high radial-velocity precision ($<5-10$ m/s) for large samples 
of M dwarfs (e.g., Eggenberger \& Udry 2010, and references therein), and 2) their being
considered as providers of very inhospitable environments for potentially
habitable planets (e.g., Tarter et al. 2007; Scalo et al. 2007, and references therein). 
These two paradigms are now shifting. First, the application of the transit technique to M dwarfs presents
several exciting opportunities, and the advantages are especially compelling
for the detection of transiting habitable, rocky planets. These include, for
example, improved observing windows due to the short periods of potential
planets in the stellar habitable zone (the range of distances from a given
star for which water could be found in liquid form on a planetary surface. 
E.g., Kasting et al. 1993), or the possibility to reach detection of rocky planets due to the small radii of
M dwarfs, leading to deep transits ($\Delta mag\sim0.005$ mag) easily detectable from
the ground with modest-size telescopes ($30-50$ cm class), and readily confirmable 
with present-day precision RV measurements (owing to their moderately large RV amplitudes, 
on the order of 5-10 m/s). Second, while not all concerns about their habitability have been resolved yet, 
there has been a recent change in view for planets orbiting low-mass M stars, now 
often considered as potentially hospitable worlds for life and its remote detection 
(e.g., Seager \& Deming 2010; Barnes et al. 2011, and references therein). 

Advancements in our knowledge of the complex processes of planet formation and evolution 
cannot be achieved without a detailed understanding of the role of the central star 
(through its properties such as mass and metal abundance) and its environment (the circumstellar 
disk within which the planetary population must form). For example, 
the theoretical expectations (within the framework of the standard core accretion model) 
that giant planet frequency and upper mass limits ought to be direct functions of 
stellar mass $M_\star$ and metallicity [Fe/H] (e.g., Laughlin et al. 2004; Ida \& Lin 2004, 2005; 
Kennedy \& Kenyon 2008; Mordasini et al. 2009) have so far been confirmed on relatively firm statistical grounds 
only for stars (mid-F through mid-K type) with masses close to that of the Sun (Santos et al. 2004; 
Fischer \& Valenti 2005; Johnson et al. 2007; Sozzetti et al. 2009), while results for stars with masses 
wignificantly different from that of the Sun still rely on small-number statistics (e.g., Endl et al. 2006; Bonfils et al. 2007; 
Johnson et al. 2010, 2011, 2012). Similarly, the statistical significance of the 
early evidence for a relatively high frequency of low-mass planets (Neptunes and super-Earths) 
around low-mass stars (e.g., Forveille et al. 2011, and references therein) is still hampered 
by the observational bias intrinsic to long-term RV surveys (only a few hundred objects monitored), 
and the recent, compelling evidence from Kepler photometry (e.g., Howard et al. 2011) of increasing 
occurrence rates for small-radius, short-period planets around increasingly cooler stars still suffers 
from small-numbers statistics at the latest spectral types (only a few hundred of relatively bright 
M0-M1 dwarfs being included in the Kepler catalogue). 
Finally, the anticipated wild diversity of the structural and atmospheric properties of 
super-Earths (Seager \& Deming 2010, and references therein) can be most easily investigated 
using a sample of such planets observed as transiting companions to nearby M dwarf primaries, given 
that for low-mass stars the planet-to-star flux ratio is much larger than that for the Earth-Sun system
\footnote{For example, in the Rayleigh-Jeans limit, this flux ratio depends on the relative surface 
areas and brightness temperatures of the planet and star. For a 2-$M_\oplus$ super-Earth, this ratio is 
in the range 0.01\%-011\% for a mid- to late-M dwarf primary (M4V-M8V), compared to 0.00044\% for 
the Earth-Sun system.}, thus spectral characterization of the planet via, e.g., occultation spectroscopy 
is much more readily attainable. 

These considerations have brought about renewed efforts to monitor
photometrically as well as spectroscopically large samples of nearby cool
dwarfs. The first spectacular success of the dedicated MEarth transit search
for rocky planets around 2000 late M dwarfs was announced by Charbonneau et al. (2009), with the detection 
of the low-density transiting super-Earth GJ 1214b ($M_p=6.5$ M$_\oplus$, $R_p=2.7$ $R_\oplus$) around 
a nearby M4.5 dwarf. The primary in this system is bright enough to enable the detailed spectroscopic characterization of the
planet's thick atmosphere over a broad wavelength range (Bean et al. 2010, 2011; Croll et al. 2011; Crossfield et al. 2011). 
The recent constraints on GJ 1214b's atmospheric composition are not only essential for breaking the 
degeneracy between the mass, radius and composition of both the interior and a possible atmosphere in 
theoretical models of super-Earths (Adams et al. 2008; Rogers \& Seager 2010; Miller-Ricci \& Fortney 2010; 
Miller-Ricci et al. 2011; D\'esert et al. 2011; Nettelmann et al. 2011; Menou 2011), but they also constitute 
a remarkable test of planetary evolution models in a mass range (for both the primary and the planet!) 
not seen in our Solar System. Very recently, the M2K Doppler search for close-in planets around 1600 nearby
M and K dwarfs has also started producing its first results (e.g., Apps et al. 2010). Decade-long Doppler 
monitoring has also allowed to detect the first Saturn-mass planet in the habitable zone of a nearby mid-M dwarf
(Haghighipour et al. 2010). The early-M dwarf GJ 581, already hosting a system 
of four low-mass (Neptunes and super-Earths) planets, is currently the focus of 
a hot debate on the actual existence of a fifth planet with the mass of a super-Earth orbiting right in the 
middle of the habitable zone (Vogt et al. 2010; Tuomi 2011; Pepe et al. 2011; Gregory 2011). 
There is a growing consensus among the astronomers' community that the first habitable rocky planet will be discovered (and might have 
been discovered already!) around a red M dwarf in the backyard of our Solar System. 

However, not all physical properties of low-mass stars are known precisely enough for the purpose of the 
detection and characterization of small-radius planets. Worse still, some of the characteristics intrinsic to 
late-type dwarfs can constitute a significant source of confusion in the interpretation in planet detection 
and characterization measurements across a range of techniques. First of all, there exist 
discrepancies between theory and observations in the determination of the sizes of M dwarfs, 
typically on the order of 10\%-15\% (Ribas 2006; Beatty et al. 2007; Charbonneau et al. 2009, and references therein). 
It has been suggested that this problems might be stemming from the lack of a detailed treatment of 
the effects of non-zero magnetic fields on the properties of low-mass, fully convective stars 
(Ribas 2006; L\'opez-Morales 2007; Torres et al. 2010, and references therein). As a result, the inferred 
composition of a transiting planet detected around an M dwarf might be subject to rather large uncertainties, 
particularly when it comes to super-Earths, for which, as mentioned above, degeneracies in the models of their physical 
structure indicate a wide range of possible compositions for similar masses and radii 
(Seager \& Deming 2010, and references therein). Indeed, for the two known transiting planets around M dwarfs, 
GJ 436b and GJ 1214b, uncertainties in the planetary parameters are dominated by the limits in the knowledge of 
the stellar parameters. 

Second, there are at present difficulties in spectroscopically determining with a high degree of precision 
M dwarf metallicities\footnote{M-dwarf spectra are dominated by chemically complex molecular features. 
As a result, the identification of the continuum in an M dwarf spectrum is challenging, rendering line-based 
metallicity indicators unreliable. The poorly constrained molecular opacity data currently available make the determination
of metallicity through spectral synthesis also difficult (e.g., Gustafsson 1989; West et al. 2011, and references therein).} 
(Bean et al. 2006; Woolf \& Wallerstein 2006; Woolf et al. 2009; Rojas-Ayala et al. 2010), 
which are only partially mitigated by recent attempts at deriving photometric calibrations 
(Bonfils et al. 2005; Casagrande et al. 2008; Johnson \& Apps 2009; Schlaufman \& Laughlin 2010). In addition, studies of the 
rotation-activity relation\footnote{The connection between stellar rotation and activity is usually 
investigated by means of 1) spectroscopic measurements of the rotational velocity $v\sin i$, usually coupled to 
measurements of the H$\alpha$ luminosity (e.g., Reiners \& Basri 2010, and references therein), 
2) spectroscopic monitoring of temporal evolution of the $R^\prime_{HK}$ activity index 
as determined from the Ca II H \& K emission line cores (e.g., Wright et al. 2004), 
and 3) photometric determination of rotation periods for stars with significant spot coverage (e.g., Strassmeier et al. 2000).} 
for M dwarfs using large stellar samples are limited to young and active stellar samples (e.g., Shkolnik et al. 2009; 
L\'opez-Santiago et al. 2010, and references therein), often in 
young open clusters (e.g., Meibom et al. 2009; Hartman et al. 2009, and references therein), 
while our understanding of the rotation-activity connection for M dwarfs with age greater than $t\sim 0.5$ Gyr 
(e.g., Pizzolato et al. 2003; Reiners 2007; Jenkins et al. 2009; West \& Basri 2009) is still subject to rather large uncertainties 
due to the sparseness of the data. 
All these issues hamper at present the possibility of determining precisely the ages of (particularly mid- and late-) M dwarfs in the field, 
and this in turn has a significant impact on the calibration of the fundamental evolutionary properties of the planets they might be hosting. 

Third, as measurements of chromospheric activity indicators (H$\alpha$ line) have shown how 
the fraction of active M dwarfs increases as a function of spectral sub-type (e.g., Bochanski et al. 2005; 
West et al. 2011), activity-related phenomena such as stellar spots, plages, and flares become increasingly a matter 
of concern for planet detection and characterization programs targeting late-type stars. Stellar surface inhomogeneities 
can hamper the detection, and sometimes even mimic the signal, of exoplanets (e.g., Queloz et al. 2001), 
and seriously complicate the characterization of their properties. This problem has already 
become acute in the case of active solar analogs hosting transiting planets. An illustrative example is provided by 
the ongoing debate on the actual mass of CoRoT-7b, varying (including 1-$\sigma$ uncertainties as large as 20\%) 
between 1 $M_\oplus$ and 9 $M_\oplus$ (!), depending on how one decides to deal with the modelling of the 
planetary signal superposed to the much larger activity-induced stellar `jitter' in both the photometric and the 
radial-velocity measurements (Queloz et al. 2009; Hatzes et al. 2010; Pont et al. 2011; Ferraz-Mello et al. 2011). 
Recently, the first serious studies attempting to gauge the limits to planet detection induced by stellar 
activity-related phenomena, and strategies aiming at minimizing such effects, have been undertaken. 
These have focused primarily on the impact of, and possibility of calibrating out, activity-induced 
jitter in high-precision radial-velocity and astrometric measurements (Makarov et al. 2009; Lagrange et al. 2010; Boisse et al. 2011, 
and references therein; Dumusque et al. 2011; Sozzetti 2011, and references therein). 
Very recently, the first analyses of the 
impact of starspots on radial-velocity searches for earth-mass planets in orbit about M dwarf
stars have been carried out by Reiners et al. (2010) and Barnes et al. (2011), who also addressed the merit of moving from the optical 
to infrared wavelengths (where the starpots-induced RV noise might be significantly reduced). 

All the above considerations clearly underline how achieving the goal of the detection {\it and} characterization of low-mass, potentially habitable, 
rocky planets around low-mass stars requires the construction of a large (all-sky) sample of nearby, relatively bright M dwarfs with 
well-characterized properties. This will necessitate the combined use of time-series of spectroscopic, astrometric, and photometric 
data of high quality. In particular, the jitter levels will have to be quantified in detail for each target individually, 
as the jitter properties may vary from star to star within the same spectral class, as suggested by recent findings 
based on high-precision Kepler photometry (e.g., Ciardi et al. 2011) and high-resolution, high-S/N spectroscopy (e.g., Zechmeister et al. 2009). 
With kick-off in December 2009, we have carried out a year-long pilot study for an 
upcoming photometric transiting search for small-size planets 
around thousands of nearby M dwarfs which will utilize an array of five 40-cm telescopes 
at the Astronomical Observatory of the Autonomous Region of the Aosta Valley (OAVdA), in the
western Italian Alps. The OAVdA site was selected on the basis of a detailed
site characterization study (Damasso et al. 2010).
The pilot study was focused on the medium-term (typically for 2 months) photometric monitoring, 
using small-size instrumentation (25-80 cm class telescope systems), of a sample of $23$ cool M0-M6 
dwarfs with good parallaxes from the TOrino Parallax Program (TOPP; Smart et al. 2010). 
The primary objectives we set out to achieve in this study were $a)$ to demonstrate
sensitivity to $<4$ R$_\oplus$ (i.e., smaller than radius of Neptune) transiting planets with periods of a few days around our
sample, through a two-fold approach that combines a characterization of the 
statistical noise properties of our photometry with the determination of transit detection 
probabilities via simulations, and $b)$ where possible, to better our knowledge 
of some astrophysical properties (e.g., activity, rotation, age) of our targets through a combination of 
spectroscopic and astrometric information and our differential photometric measurements. 

In \S~2 we describe the OAVdA instrumentation utilized during the pilot study, and outline 
the dedicated pipeline for the data processing and analysis of the photometric data we have developed. 
We discuss in \S~3 the main characteristics of the M dwarf sample targeted by the pilot study, and 
present in \S~4 the main results of the study in terms of 1) achieved short-term and medium-term photometric sensitivity 
for our sample, 2) improved characterization of the properties of the cool M dwarfs observed, by combining 
the knowledge of their photometric micro-variability time-scales with other available spectroscopic, 
photometric, and astrometric observations, and 3) a careful assessment of the limits to transiting 
planetary companions for each star in our sample. We conclude in \S~5 by 
summarizing our findings and by discussing the preparatory steps for the upcoming 
long-term photometric monitoring program to characterize the micro-variability features 
of and search for transiting small-size planetary companions to a well-defined sample of low-mass stars, 
to be carried at the OAVdA site in the near future.

\section{Instrumentation and methodology}\label{sec:metod}

The instrumental set up used in this study is very similar to the one adopted for the OAVdA site characterization study 
described in Damasso et al. (2010). In summary, we used a small-size telescope array composed of 
three instruments with diameters of 81 cm, 40 cm, and 25 cm, respectively. For the purpose of the pilot study described here, 
the 40-cm  and  the 25-cm telescopes performed all the observations equipped with standard Johnson-Cousins $I$ filters, 
while the 81-cm utilized a standard Johnson-Cousins $R$ filter (observations in the $I$ filter with this instrument were affected 
by fringing). Naturally, the choice of filters was driven by red colours of the target sample. 
In Table~\ref{table:1} we summarize the main characteristics of the telescope and camera systems.

\begin{table*}
\caption{Summary of the main characteristics of the telescope and camera systems.}             
\label{table:1}      
\centering                          
\begin{tabular}{c c c c c c c}        
\hline\hline                 
&\multicolumn{2}{c}{Telescope}&\multicolumn{2}{c}{CCD camera}&\multicolumn{2}{c}{Resulting configuration} \\  
\hline
Optical scheme&Aperture&Focal ratio&Sensor area&Pixel area&FoV&Plate scale\\
&(cm)& &(pixel$^{2}$)&($\mu$ m$^{2}$)&(arcmin$^{2}$)&($^{\prime\prime}$/pixel)\\
\hline                   
Reflector Maksutov & 25 & f/3.80 & $2184\times1472$ & $6.8\times6.8$ & $52.10\times35.11$ & 1.43 (binning $1\times1$) \\
Reflector Ritchey-Chr\'{e}tien & 40 & f/7.64 & $1024\times1024$ & $24\times24$ & $26.4\times26.4$ & 1.55 (binning $1\times1$) \\
Reflector Ritchey-Chr\'{e}tien & 81 & f/7.90 & $2048\times2048$ & $15\times15$ & $16.3 \times16.3$ & 1 (binning $2\times2$)\\
\hline\hline                            
\end{tabular}
\end{table*}

All observations were performed with the CCDs set up at the focus of the telescope, and the exposure times were chosen 
at the beginning of each night of observation 
(without being subsequently modified) in order to guarantee an optimum signal-to-noise ratio (S/N$\gtrsim100$) for the target, 
while avoiding saturation (see Table~\ref{table:2} for the typical exposure times for each monitored target). 
The temporal sampling, including overheads, was typically of a few minutes. We chose to observe with any 
given instrument one star at a time during a night, tracking it without repointing. No auto-guiding was utilized for 
the 25cm and 40cm telescopes (as described already in Damasso et al. 2010). For the two instruments, we 
recorded typical drifts of up to ~100 pixels. None of the above elements of the observing strategy was optimized 
with the intent of reproducing the actual one to be implemented in our upcoming survey, which will employ vastly 
improved isntrumentation and an adaptive observing strategy to be described in a future work. In this respect, 
the resulting photometric performances reported here can be considered as conservative. 

For the pilot study we selected $23$ targets from the input target list of the TOPP Program (see \S~\ref{sec:stellar} for details). 
We chose an observational strategy which would allow us to monitor each target for at least 3 hr/night
without interruptions for a minimum of a dozen nights over a maximum period of $\sim2$ months. The expected 
phase coverage for transiting companions within a few (1-2) days of period would exceed $70\%$. Two stars, 
LHS 1976 and LHS 534, were observed for a significantly longer period of time (LHS 1976 actually over two seasons), 
in order to probe our sensitivity to transiting companions on periods of up to about 1 week or so. 
At the end of the pilot study (a little over 1 year of observations) the whole database comprised 76287 good images, 
corresponding to $\sim 1000$ GB of data. A summary of the pilot study observations is provided in Table~\ref{table:2}. 

\begin{table*}
\caption{Log of observations for the target sample.}             
\label{table:2}      
\centering                          
\begin{tabular}{c c c c c c c}        
\hline\hline                 
ID&LHS&Number of frames&Number of nights&Epoch range&Telescope&Typical exposure\\
&&&&(days)&(cm)&(sec)\\
\hline        
1&1104&2049&14&48&40&61\\
2&1475&6826&20&72&25&34\\
3&228&1084&18&54&40&155\\
4&243&774&14&44&40&137\\
5&1976&24761&108&482&25/81&64/7\\
6&6158&2804&10&40&25&28\\
7&269&1629&14&56&40&60\\
8&2220&3980&16&58&40&23\\
9&283&1696&12&56&25&58\\
10&2472&1022&8&57&25&75\\
11&360&46&1&1&40&120\\
12&\dots&4421&15&64&25&21\\
13&417&4693&34&85&25/81&10/3\\
14&3343&2122&17&88&25/40&60/30\\
15&3445&2835&19&66&25&54\\
16&528&1697&13&40&25&62\\
17&534&7346&44&122&81&8\\
18&1721&3028&22&61&81&6\\
19&370&633&9&33&81&120\\
20&306&75&1&1&40&75\\
21&\dots&1265&11&59&25&75\\
22&2686&914&10&60&40&75\\
23&2719&587&7&20&40&90\\
\hline\hline                            
\end{tabular}
\end{table*}

The data reduction procedure utilizes an uprgaded version of TEEPEE (Transiting ExoplanEts PipElinE), 
described in detail in Damasso et al. (2010). In short, TEEPEE is a software package written in 
IDL\footnote{IDL is a commercial programming language and environment by ITT Visual Information Solutions. 
http://www.ittvis.com/idl/}, which utilizes publicly available software from the Astronomy Users' Library
\footnote{http://idlastro.gsfc.nasa.gov/contents.html} as well as external contributed FORTRAN routines. 
The areas of the software written from scratch by two of us (PG and MD) are the ones devoted to automatically 
carry out ensemble differential aperture photometry on an user-specified target. TEEPEE is organized in three main, 
sequential modules:
\begin{itemize}
\item image calibration (including dark and bias subtraction, and flat fielding)
\item astrometric processing (image alignment) and photometric processing (aperture photometry)
\item differential photometry of the target with respect to a chosen set of reference stars
\end{itemize}

The heart of TEEPEE, for the purpose of this study, is the third block. 
It performs those operations which are necessary to correct, to a high degree of reliability, for systematic effects 
which cause the degradation of the precision of the photometric measurements, and it produces photometric 
light curves for hundreds of stars detected in the field. Our differential photometric method performs as follows. 
For each frame \textit{i}, we use as the reference magnitude \textit{M$_{rif}^{i}$} the average 
magnitude of the \textit{n} reference stars:

\begin{equation}
\centering
M_{rif}^{i}=\frac{\sum_{k=0}^{n} M_{k}^{i}}{n}. 
\end{equation}

\textit{M$_{rif}^{i}$} is then subtracted to the magnitude of the user-defined target 
\textit{M$_{target}^{i}$}, obtaining the difference $\Delta M^{i}=M_{rif}^{i} - M_{target}^{i}$. 
The procedure is iteratively repeated for all the reference stars, using as new references 
the remaining $n-1$ stars. This procedure is also iteratively repeated for all detected field stars 
(based on a 3-$\sigma$ above background criterion), using as references 
the $n$ reference stars chosen for the target.

The second and the third modules have been upgraded with respect to the previous version of the software described 
in Damasso et al. (2010) in order to improve the differential aperture photometric processing in an automatic way.

In the second module we implemented a multi-aperture photometric processing. We settled on twelve apertures, 
typically ranging between 2 and 4 times average FWHM (varying slightly depending on the telescope used).  

In the third module, we first take care of picking up reference objects on a CCD sub-frame, avoiding the chip edges, 
affected by vignetting which is not fully corrected for during flat-fielding. 
Second, we use two efficient methods, based on the Burke et al. (2006) prescription, for choosing the appropriate 
set of references for the target. The first method ($m1$) selects the subset of reference stars which minimizes 
the RMS of the differential light curve of the target; the second method ($m2$) selects the subset of references 
which minimizes the RMS of the differential light curve of each potential reference star. Both methods are then applied 
to all 12 apertures in order to choose the optimal one, on the basis of a minimum-RMS prescription for the target light curve.
Next, we filter out outliers using a 3-$\sigma$ clipping criterion, and an additional light curve correction to 
suppress FWHM-related effects (e.g., Irwin et al. 2007).
Finally, we apply the SysRem trend filtering algorithm (Tamuz et al. 2005), to correct for unknown systematic 
effects in the light curves produced with methods $m1$ and $m2$. The two resulting light curves are dubbed as produced 
with methods $m3$ and $m4$, respectively. The whole procedure thus produces a total of four light curves for each target, 
for each night.

With a similar process, four additional light curves are produced for each target over the whole timespan of the observations, 
which we dub `full-period' light curves. A single (the more stable) set of reference stars is chosen over the entire observation window. 
The only information retained from the nightly procedure is the optimal aperture for each night. In place of SysRem, a 
different trend filtering algorithm is utilized, TFA (Kov\'acs et al. 2005), which after some experimenting was found 
to be better suited for the full-period light curve analysis.

\section{Stellar sample}\label{sec:stellar}

As already mentioned earlier, the M dwarfs targeted during the pilot study were selected from the TOPP program input list 
of nearby ($d< 60$ pc) cool stars. 
We chose stars spanning a range of spectral sub-types, approximately M0 through M6, and set an $I$-band magnitude limit of $I\lesssim14$
(thus keeping exposure times typically to within a few minutes). 
These two criteria allowed us to probe potentially different regimes of intrinsic stellar variability (see \S~\ref{intro}), both in terms of amplitude and time-scales.
 
Most of these objects, with the exception of LHS 228, LHS 306, LHS 360, and LHS 2719, are included in the LSPM-North catalog 
(L\'epine \& Shara 2005), with either Hipparcos trigonometric parallaxes or distance moduli estimates indicating they are within 33 pc 
from the Sun (L\'epine 2005). Furthermore, for about half of the sample Smart et al. (2010) have published precision astrometric 
information, including direct distance estimates. 
For the purpose of this study, and particularly to derive the results presented 
in \S~\ref{limits}, it is important to provide reliable estimate of mass $M_\star$ and radius $R_\star$ for all our program stars.
For the targets presented in Smart et al. (2010), we utilized the values of $M_\star$ therein. 
For the other targets, we derived absolute $K$ magnitudes from the distance estimates
and we obtained $M_\star$ from the mass-luminosity calibration of Delfosse et al. (2000). For all the program stars, 
the Bayless \& Orosz (2006) mass-radius relation was then employed to calculate $R_\star$. For completeness, metallicity 
[Fe/H] estimates are provided using the Johnson \& Apps (2009) calibration in the plane $V-K$-$M_{K}$, where applicable 
and only for objects with trigonometric parallaxes, along with effective temperature 
$T_\mathrm{eff}$ values from the Casagrande et al. (2008) M dwarfs ($V-K$)-$T_\mathrm{eff}$ calibration. All the above characteristics of our 
sample are summarized in Table~\ref{table:3}. For each target, the same Table also reports relevant information on 
flaring activity (obtained from SIMBAD) and activity indicators, i.e., H$\alpha$ equivalent widths measurements.

\setlength{\tabcolsep}{0.04in}
\ctable[
caption = Characteristics of the M dwarf sample.,
center,
star,
label=table:3,
doinside = \footnotesize]
{r l c c c c c c c c c c c c c c c}
{
\tnote[a]{Trigonometric parallax / V magnitude after Smart et al. 2010 and references therein}
\tnote[b]{Gizis, Reid \& Hawley 2002} 
\tnote[c]{Trigonometric parallax / V magnitude after Henry et al. 2006} 
\tnote[d]{Walkowicz \& Hawley 2009}
\tnote[e]{Photometric parallax / V magnitude after L\'epine \& Shara 2005}
\tnote[f]{Trigonometric parallax from Hipparcos data, after Van Leeuwen 2007} 
\tnote[g]{V magnitude from ASCC-2.5 catalogue, after Kharchenko 2001} 
\tnote[h]{Trigonometric parallax / V magnitude after Reid \& Cruz 2002} 
\tnote[i]{Shkolnik et al. 2009} 
\tnote[l]{TASS Mark IV photometric catalogue, version 2, after Droege et al. 2006}
}{
\FL
ID& LHS&GJ&RA&DEC&$d$&$V$&$J$&$K$&$M_V$&$M_K$&$M_\star$&$R_\star$&$T_\mathrm{eff}$&[Fe/H]&Flare&H$\alpha$ EW \NN
&&&(hh:mm:ss)&(dd:mm:ss)&(pc)&&&&&&($M_\odot$)&($R_\odot$)&(K)&(dex)&star&(\AA)\ML
1 &  1104 &\dots&00:35:53.1&+52:41:13.7&24.7$\pm$1.3\tmark[a]&12.54\tmark[a]& 8.93&8.09&10.58&6.13&0.44&0.45 &3317& 0.07&n&\dots \NN 
2 &  1475 & 119 &02:56:34.4&+55:26:14.0&16.7$\pm$1.4\tmark[a]&10.48\tmark[a]& 7.43&6.59&9.37&5.48&0.56&0.57  &3620&\dots&n&-0.534\tmark[b]\NN 
3 &  228  &\dots&07:16:27.7&+23:42:10.4&42.4$\pm$2.0\tmark[a]&15.53\tmark[a]&12.02&11.30&12.39&8.17&0.12&0.14&3424&-1.27&n&\dots\NN 
4 &  243  &\dots&08:03:06.1&+34:56:54.8&24.4$\pm$0.7\tmark[a]&16.09\tmark[a]&11.51&10.74&14.15&8.81&0.09&0.11&2986&-0.67&n&\dots\NN
5 &  1976 &\dots&08:03:19.5&+52:50:38.1&24.8$\pm$3.1\tmark[a]&11.38\tmark[a]& 8.06&7.24&9.407&5.26&0.58&0.58 &3472& 0.29&n&\dots\NN
6 &  6158 &3522 &08:58:56.3&+08:28:25.9& 6.8$\pm$0.1\tmark[c]&10.92\tmark[c]& 6.51&5.69&11.76&6.53&0.36&0.37 &3023& 0.51&y&+2.76\tmark[d]\NN
7 &  269  &\dots&09:29:11.0&+25:58:09.3&16.7$\pm$1.3\tmark[e]&16.43\tmark[h]&10.91&9.96&15.32&8.85&0.11&0.14 &2697&\dots&n&\dots\NN 
8 &  2220 &3585 &10:06:43.8&+41:42:52.6&20.7$\pm$1.2\tmark[f]&11.31\tmark[g]& 8.21&7.40& 9.73&5.82&0.49&0.50 &3607&-0.21&n&\dots \NN
9 &   283 &3612 &10:35:26.9&+69:26:58.8&13.2$\pm$0.7\tmark[h]&11.95\tmark[h]& 7.90&7.16&11.35&6.56&0.35&0.37 &3175& 0.13&n& -0.219\tmark[b], +0.13\tmark[d]\NN 
10 &  2472&452.1&11:54:07.9&+09:48:22.8&11.3$\pm$0.5\tmark[a]&12.81\tmark[a]& 8.70&7.87&12.54&7.60&0.20&0.22 &3120&-0.33&y&+1.248\tmark[b]\NN
11 &  360 &\dots&13:46:55.5&+05:42:56.3&56.2$\pm$9.5\tmark[a]&15.22\tmark[a]&12.39&11.66&11.47&7.87&0.28&0.29&3858&\dots&n&\dots\NN 
12 &\dots &569B &14:54:29.2&+16:06:03.8& 9.7$\pm$0.2\tmark[f]&10.38\tmark[g]& 6.63&5.77&10.45&5.84&0.48&0.49 &3247& 0.37&y&+0.852\tmark[b]\NN
13 &  417 & 623 &16:24:09.3&+48:21:10.4& 8.1$\pm$0.08\tmark[f]&10.27\tmark[g]& 6.64&5.91&10.73&6.37&0.38&0.40&3359&-0.14&n&-0.237\tmark[b]\NN
14 &  3343&4040 &17:57:50.9&+46:35:19.1&14.6$\pm$0.5\tmark[a]&11.68\tmark[a]& 7.85&7.00&10.85&6.17&0.40&0.41 &3219& 0.25&n&-0.215\tmark[b], +0.15\tmark[d]\NN
15 &  3445&9652A&19:14:39.1&+19:19:03.7&17.9$\pm$0.7\tmark[a]&11.59\tmark[a]& 7.58&6.81&10.32&5.54&0.41&0.42 &3179& 0.69&y&+5.4\tmark[d], +5.7\tmark[i]\NN 
16 &  528 &1271 &22:42:38.7&+17:40:09.2&17.4$\pm$0.7\tmark[a]&11.76\tmark[a]& 8.06&7.18&10.55&5.98&0.43&0.44 &3260& 0.26&n&\dots\NN
17 &  534 &4311 &23:06:35.6&+71:43:25.5&17.3$\pm$5.2\tmark[e]&11.75\tmark[l]& 8.34&7.56&10.56&6.37&0.37&0.39 &3445& \dots &n&\dots\NN 
18 &  1721&1074 &04:58:46.0&+50:56:37.8&19.3$\pm$0.9\tmark[f]&11.06\tmark[l]& 7.90& 7.04&9.63&5.61&0.53&0.54 &3540&-0.01&n&-0.452\tmark[b]\NN 
19 &  370 &\dots&14:20:53.1&+36:57:16.8&21.7$\pm$1.7\tmark[e]&16.17\tmark[h]&11.05&10.25&14.49&8.57&0.13&0.16&2829& \dots &n&\dots\NN 
20 &  306 &3668 &11:31:08.3&-14:57:21.1&11.8$\pm$2.6\tmark[a]&14.19\tmark[a]& 9.36&8.50&13.83&8.14&0.12&0.14 &2889&-0.05&n&\dots\NN 
21 &\dots &1167A&13:09:34.9&+28:59:06.6&10.8$\pm$0.6\tmark[a]&14.52\tmark[a]& 9.48&8.61&14.35&8.44&0.12&0.14 &2832&-0.08&y&+4.82\tmark[i]\NN 
22 &  2686&\dots&13:10:12.6&+47:45:18.4&10.3$\pm$0.5\tmark[a]&14.16\tmark[a]& 9.58&8.69&14.09&8.62&0.09&0.11 &2951&-0.47&n&\dots\NN 
23 &  2719&\dots&13:20:27.0&-03:56:14.5&26.9$\pm$2.5\tmark[a]&15.60\tmark[a]&11.28&10.48&13.45&8.33&0.15&0.17&3058&-0.58&n&\dots\LL                          
}

\section[]{Results}

\subsection{Photometric precision}\label{sec:photprec}

Accurately gauging the short- and medium-term photometric precision for our red dwarfs sample, 
as a function of the characteristics of the instruments adopted and of the details of the 
methods of differential photometry utilized in the analysis, and including meaningful estimates 
of the degree of correlated (red) noise in the data, is of fundamental importance. The short-term 
precision (as defined below) is the quantity that ultimately has the most relevant impact 
on the probability of detecting a transit signal, and of determining its statistical significance. 
The medium-term precision is the quantity that allows one to gather insight on different timescales 
of variability which are more likely to be intrinsic to the target, such as surface activity. 
As is usual practice, we operationally define the photometric precision of a given target 
monitored with one of our instruments as the RMS of a differential light curve (for one observing night or 
for the whole time interval of the observations) obtained with any of the four methods ($m1$ through $m4$) defined in \S~\ref{sec:metod}. 
To first order, this corresponds to the uncertainty of each data point. 

\subsubsection{Global analysis}\label{phot_prec}

Our first goal is to determine the precision of our photometric measurements on a nightly basis (short-term precision) and over the 
whole temporal length of the observations (medium-term precision). 
For each target, we determine the short-term photometric precision of the data by computing the intra-night RMS values of eight light curves: 
four derived from the nightly light curves and four extracted from the full-period light curves. 
We determine the medium-term precision in our dataset by computing the global RMS values of the four full-period light curves. 

For the purpose of computing the RMS values, we utilized unbinned light curves for the 25-cm and 40-cm telescopes, 
while we utilized binned light curves (at 1.5 min) for the 81-cm telescope. The binning of the 81-cm telescope 
data is necessary in order to homogenize them with the typical time sampling of the other two telescopes, and more 
importantly to suppress the effects of atmospheric scintillation, which usually affects to a significant degree 
very short (few seconds) exposures.  

\begin{figure}
\centering
\includegraphics[width=0.45\textwidth]{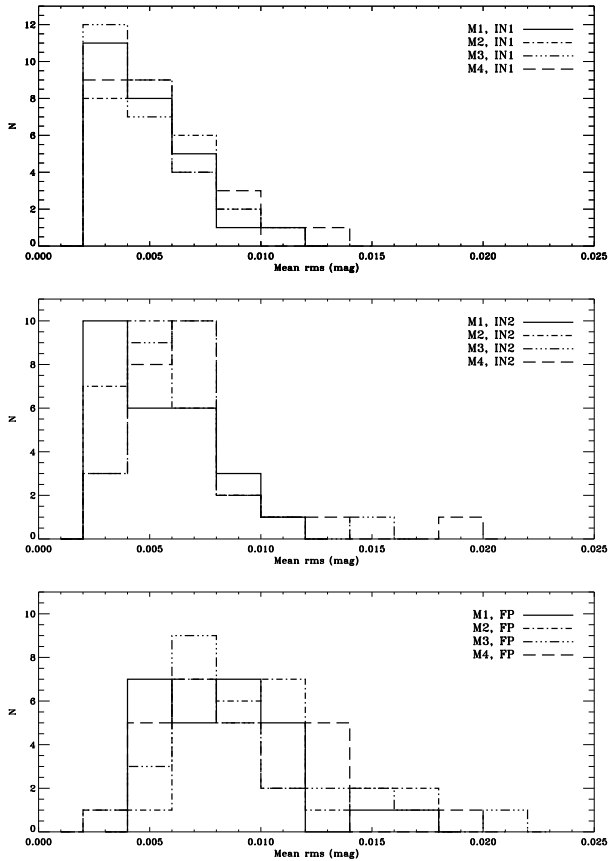}
\caption{Distribution of the mean RMS for each target. For each of the four differential photometry methods, 
the panels show: the mean intra-night RMS distribution IN1 for the sample (top), the mean intra-night RMS 
distribution IN2 extracted from a full-period light curve for each target (center), 
and the full-period RMS distribution FP for the sample (bottom).} 
\label{figpre}
\end{figure}

In the three panels of Figure~\ref{figpre} we show the distributions of the mean short- and medium-term 
RMS values for each target, for the various methods.
Visual inspection of the plots already shows qualitatively two relevant results: the apparently good agreement 
between different methods for differential photometry (with or without detrending), both in the short- and medium-term 
regimes, and the fact that the medium-term photometric precision appears degraded with respect to the short-term 
precision. We put the discussion on more firm statistical grounds by performing a comparison between the various RMS 
distributions based on the Kolmogorov-Smirnov (K-S) test, which measures the probability $P_{K-S}$ of any two distributions being 
statistically indistinguishable. 

First, we perform the K-S test between the four distributions within each panel of Figure~\ref{figpre}, to 
gauge statistical differences in the RMS distributions between the $m1$ through $m4$ analysis methods, 
separately for the short- and medium-term regimes. The results, expressed in terms of $P_{K-S}$, are summarized in Table~\ref{table:ks1}. 

\begin{table}
\caption{K-S test results (1). }             
\label{table:ks1}      
\centering                          
\begin{tabular}{c c c c c c}        
\hline\hline                 
m1-m2&m1-m3&m1-m4&m2-m3&m2-m4&m3-m4\\
\hline 
&&\multicolumn{2}{c}{Figure~\ref{figpre}, upper panel}&&\\
\hline        
0.992&0.992&0.992&0.674&0.9928&0.674\\
\hline 
&&\multicolumn{2}{c}{Figure~\ref{figpre}, middle panel}&&\\
\hline 
0.992&0.138&0.258&0.440&0.258&0.999\\
\hline 
&&\multicolumn{2}{c}{Figure~\ref{figpre}, lower panel}&&\\
\hline 
0.138&0.674&0.674&0.440&0.258&0.992\\
\hline\hline                            
\end{tabular}
\end{table}

Next, the K-S test is performed to evaluate the differences between the RMS distributions in the short- 
and medium-term regimes. Given the results shown in Figure~\ref{figpre} (i.e., methods $m1$ through $m4$ provide 
the same answers), we show here only for simplicity the comparison between the RMS distributions 
of method $m1$ from one panel with all other distributions in the other panels. 
The corresponding values of $P_{K-S}$ are summarized in Table~\ref{table:ks2}. 

\begin{table}
\caption{K-S test results (2). }             
\label{table:ks2}      
\centering                          
\begin{tabular}{c c c c}        
\hline\hline                 
m1-m1&m1-m2&m1-m3&m1-m4\\
\hline 
&\multicolumn{2}{c}{upper panel (Fig.~\ref{figpre})- middle panel (Fig.~\ref{figpre})}&\\
\hline        
0.258&0.138&0.012&0.031\\
\hline 
&\multicolumn{2}{c}{upper panel (Fig.~\ref{figpre})- lower panel (Fig.~\ref{figpre})}&\\
\hline 
0.001&2e-8&4e-5&0.0002\\
\hline 
&\multicolumn{2}{c}{middle panel (Fig.~\ref{figpre})- lower panel (Fig.~\ref{figpre})}&\\
0.031&0.0002&0.031&0.013\\
\hline\hline                            
\end{tabular}
\end{table}

Finally, the K-S test is performed to evaluate the differences between the RMS distributions 
from data obtained with different telescopes. The corresponding values of $P_{K-S}$ are summarized in Table~\ref{table:ks3}. 

\begin{table}
\caption{K-S test results (3). }             
\label{table:ks3}      
\centering                          
\begin{tabular}{c c c}        
\hline\hline                 
25cm-40cm&25cm-81cm&40cm-81cm\\
\hline      
0.315&0.565&0.397\\
\hline\hline                            
\end{tabular}
\end{table}

The results from the global short-term and medium-term photometric RMS analysis based on the K-S test 
highlight the following: 

\begin{itemize}
\item[1)] the use of a range of instrumental setups does not impact significantly the 
typical performance in the photometry; 
\item[2)] the differential photometry methods (without detrending) adopted within the pipeline provide consistently similar results;
\item[3)] the similarity between the different light curve RMS values with and without detrending 
indicate that our data are not particularly affected by linear systematic trends that can be easily identified and 
removed, such as atmospheric extinction, detector efficiency, or PSF changes over the detector, for which algorithms like SysRem 
are very effective;
\item[4)] the differences between the short- and medium-term photometric precision suggest that 
correlated noise (astrophysical in nature, or not) becomes important on timescales  longer than 
those of one observing session (one night). For example, the degradation in precision seen in 
the bottom panel of Fig.~\ref{figpre} could be ascribed in part to flat-fielding errors and/or to pointing errors and telescope drifts, 
the latter effectively reducing the number of reference objects used to perform differential photometry on all the 
nights at once (we recall that an object can be used as comparison star only if it is detected in every 
frame, and the probability that a star is lost in one or more frames increases with the time span covered by the observations). 
 
\end{itemize}

In summary, the typical short-term photometric precision is $\sim5$ mmag and the typical medium-term 
photometric precision is $\sim9$ mmag. The former result for the M dwarf sample is in excellent 
agreement with the findings of Damasso et al. (2010), who had focused their analysis on a small number of 
stars more similar to the Sun.

\subsubsection{Correlated (red) noise analysis}\label{rdn}

The RMS of the whole light curve can be assumed to be the uncertainty of each data point under a fundamental hypothesis: 
the photometric measurements are assumed to be uncorrelated (white noise regime). At the millimag level and in a high-cadence 
time series this is generally untrue. There are many effects that can produce a correlated photometric measurements (red noise regime): 
changing airmass, atmospheric conditions, telescope tracking (and relative flat field errors) 
and the intrinsic variability of the targets. These effects introduce some covariance between data points. 
  
The presence of red noise can have a rather significant impact on the statistical analysis of the data. 
As an example, let us consider a light curve, with no evident variability, consisting on \textit{N} flux measurements $f_i$ in a fixed time interval, 
with uncertainties for each data point $\sigma_i$ equal to the 
RMS of the whole light curve $\sigma_0$.
We calculate the mean of the \textit{N} flux measurements $f_{mean}$.
The uncertainty on $f_{mean}$, $\sigma_{f_{mean}}$, is then the error of the mean of $f_i$.
Using the expression for the standard deviation of the mean under the assumption of white noise ($\sigma_w$), this uncertainty is:

\begin{equation}
\sigma_{f_{mean}}\equiv\sigma_{w}=\frac{\sigma_0}{\sqrt{N}} .
\label{eq:wn}
\end{equation}

We note as the uncertainty on $f_{mean}$ decreases with the square root of the number of points $N$.

The equivalent of Eq.~\ref{eq:wn} in presence of red noise is:

\begin{equation}
\sigma_{f_{mean}}\equiv\sigma_{t}=\sqrt{\sigma_w^2+\sigma_r^2} = \sqrt{\frac{\sigma_{0}^2}{N}+\frac{1}{N^2}\sum_{i\not=j}C_{ij}},
\label{eq:rn}
\end{equation}

where the $C_{ij}$'s are the covariance coefficients between the \textit{i}-th and \textit{j}-th measurements (e.g., Pont et al. 2006).

Now, we can estimated $\sigma_{t}$ from the light curves, taking in account the red noise, following the methodology 
described in Pont et al. (2006), and then determine the red noise component from Eq.~\ref{eq:rn}:
\begin{equation}
\sigma_{r}=\sqrt{\frac{1}{N^2}\sum_{i\not=j}C_{ij}}=\sqrt{\sigma_{t}^2-\frac{\sigma_{0}^2}{N}}.
\label{eq:r}
\end{equation}

\begin{figure}
\centering
\includegraphics[width=0.45\textwidth]{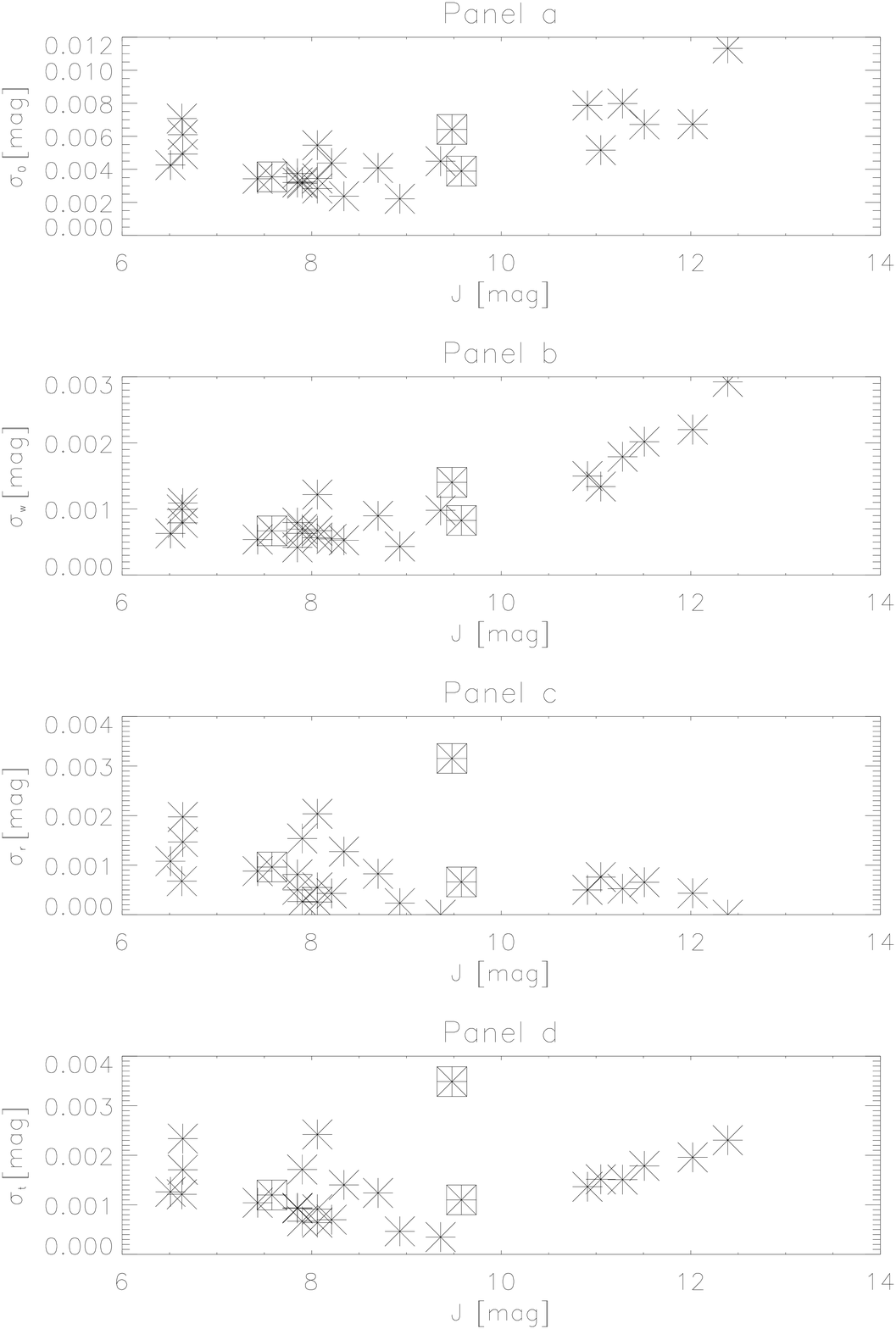}
\caption{Panel $a$: mean RMS of the unbinned light curves as a function of $J$ mag. Panel $b$: white noise term as a function of $J$ mag. 
Panel $c$: red noise term as a function of $J$ mag. Panel $d$: the total error term for the binned light curves as a function of $J$ mag. 
Stars indicated by asterisks framed by a square are objects showing clear signs of activity (see\S~\ref{sec:flare} and \S~\ref{sec:starspots})} 
\label{figred}
\end{figure}

For each star in our sample, we computed $\sigma_{t}$ over an interval $\delta t=30$ min, and repeated the calculations 
using the light curves from the four intra-night methods.
We decided to investigate the red noise contribution to the error budget over this time interval, as an illustrative example. 
Note that this interval actually corresponds to a significant fraction of the typical transit duration 
for few-days period planets orbiting mid and late M dwarfs, and thus it's of particularl interest for the purpose of this study.
We show in Figure~\ref{figred} the various contributions to the photometric error budget in our dataset, 
expressed as a function of the target $J$ mag. The results shown correspond to the error analysis for the light curves 
obtained using method $m1$. The outcome of this specific study is only marginally dependent on the chosen method for differential photometry, 
as we discuss below. 
Panel \textit{a} of Figure~\ref{figred} shows the expected trend of increasing $\sigma_{0}$ with magnitude in the data. 
Similarly, panel \textit{b} emphasizes a positive correlation of $\sigma_w$ with magnitude, obviously related to photon noise. 
This effect is also expected, as the exposure times (see \S~\ref{sec:metod}) were chosen so as to guarantee a high $S/N>100$ 
for each object, based on the theoretical noise estimate in Eq.~3 of Damasso et al. (2010).
Panel \textit{c} shows a weak inverse trend of $\sigma_r$ with $J$ mag, with decreasing red noise contribution for fainter objects. 
Again, this is expected because for faint objects $\sigma_w$ dominates the total error budget, and $\sigma_r$ becomes 
increasingly more difficult to determine at the faint end. 
Finally, panel \textit{d} shows how the total photometric error $\sigma_{t}$ in the binned data behaves as a function 
of $J$ mag due to the combination of correlated and uncorrelated noise terms.
   
\begin{figure}
\centering
\includegraphics[width=0.49\textwidth]{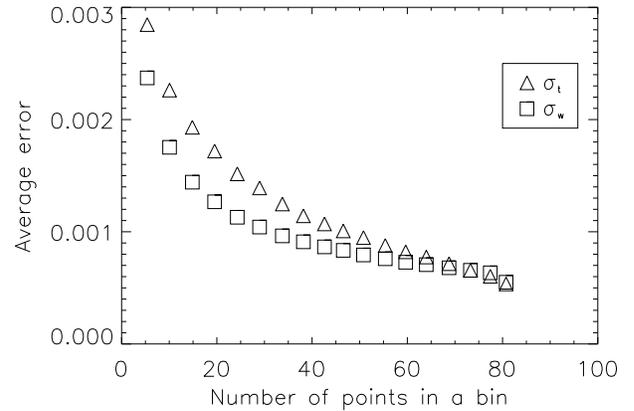}
\caption{The behaviour of $\sigma_t$ and $\sigma_w$ vs number of points in time bins} 
\label{figredt}
\end{figure}

Figure~\ref{figredt} shows, for the whole sample, the behaviour of $\sigma_t$ (triangle) and $\sigma_{w}$ (square) 
as a function of the number of flux points in a bin (or equivalently, time length of the interval $\delta t$). 
In this case the range of points corresponds to $\delta t$ ranging from 5 to 90 minutes. 
The plot highlights that the total noise is influenced by the correlation of the data points mainly on short time scale (small $N$), 
while the red component significantly decreases at longer time scales. The noise $\sigma_t$, globally, follow a $1/\sqrt{N}$ relation, 
as it is expected in a regime where the white noise is dominant. 

It is furthermore worth mentioning how the typical uncertainties over the average of the time intervals considered here, 
taking into account the red noise, are 1.4 mmag, 1.6 mmag, 1.2 mmag, and 1.4 mmag for photometric methods $m1$ through $m4$, 
respectively. This corresponds to values which are 1.3 times, 1.5 times, 1.2 times, and 1.3 times larger than a pure white noise regime. 
Based on a K-S test, such differences are deemed significant with confidence levels of 94.0\%, 99.1\%, 95.4\%, and 99.9\%, respectively.
As opposed to the results obtained in \S~\ref{phot_prec}, where we simply used the global light curve RMS as a comparison metric, 
this more in-depth analysis underlines how the detrending algorithms are actually useful for the purpose of 
suppressing some of the correlations present in our data. 

Finally, we note how the red noise analysis can also be used in principle to investigate the intrinsic 
variability of a target at the millimag level. The points framed by a square in Figure~\ref{figred} 
are the three stars in our sample showing clear signs of activity, for which we have observed either flare events 
or determined their likely photometric rotation period based on the presence of surface inhomogeneities, or both 
(see\S~\ref{sec:flare} and \S~\ref{sec:starspots}). In one case (GJ 1167A), the high levels of red noise could also 
be seen as an indication of strong activity (i.e photometric variability) at the millimag level, and on short timescales. 
Of course, in order to systematically apply this type of analysis to describe stellar activity robust calibration 
procedures on a statistically significant sample of stars are necessary, together with the help of spectroscopic 
measurements. 
This is a challenge we plan to take up in the future, 
in order to better characterize the global target sample of our upcoming survey, as this might also allow us to 
reduce the investment in spectroscopic follow-up observations.  

\subsubsection{On the choice of the comparison stars}

\begin{figure*}
\centering
\includegraphics[width=0.90\textwidth]{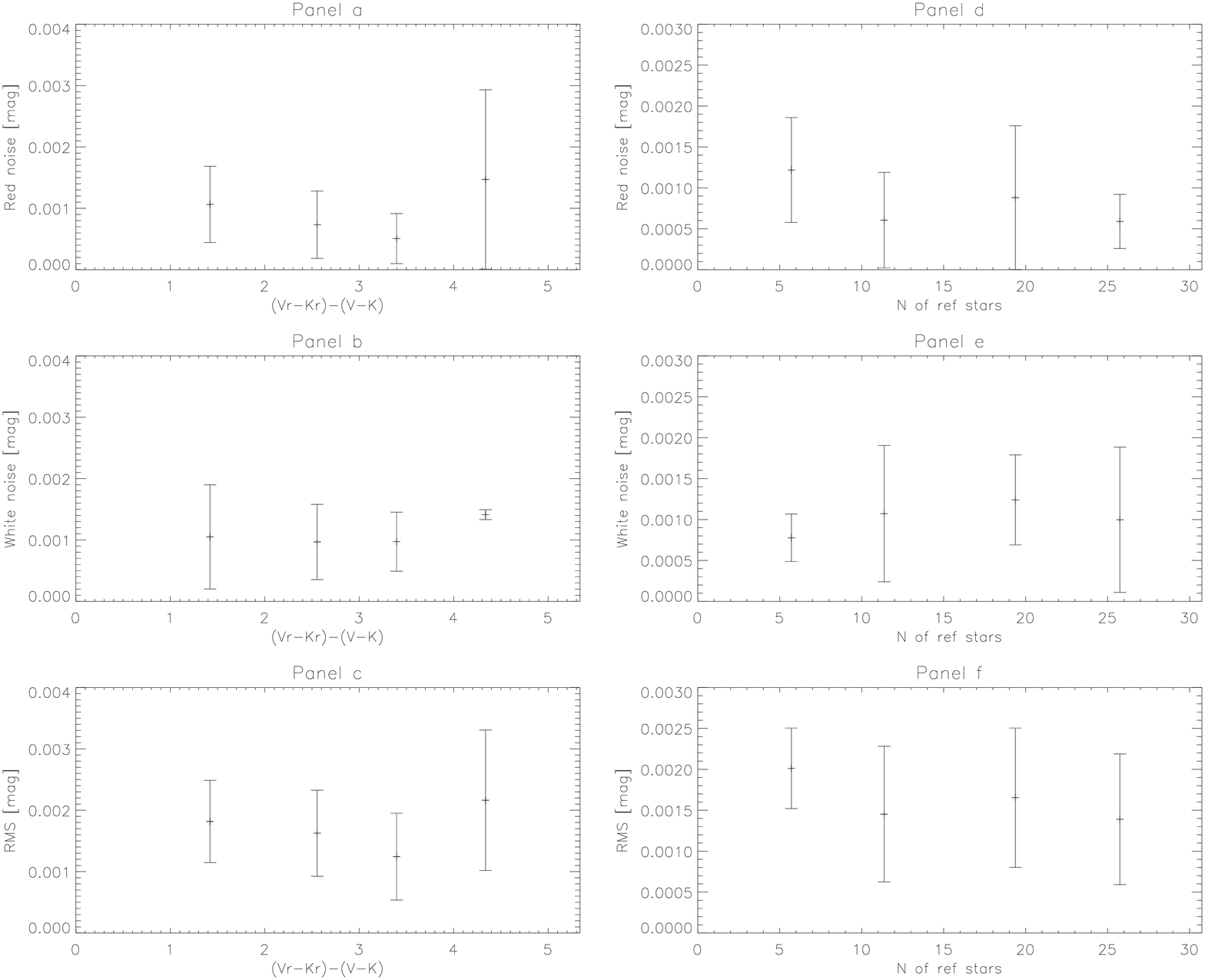}
\caption{Left: Red noise (top), white noise (middle), and total RMS (bottom) for the sample as a function of the difference $\textit{(Vr-Kr)-(V-K)}$ 
between the colour index of the target and the average color index of the corresponding comparison stars. Right: 
The same quantity is plotted as a function of the average number of comparison stars for each red dwarf in the sample.}
\label{colordep}
\end{figure*}

When characterizing the sources of correlated and uncorrelated noise in our photometric data, it is
reasonable to expect them to at least in part be due to details in the selection of the reference objects. 
This is all the more so in the case of differential photometry of late-type targets observed with red filters 
(e.g., Bailer-Jones \& Lamm 2003).
To test this possibility, for each red dwarf in our sample we calculated an `average' colour index $V-K$ of the references 
(using the magnitudes of the NOMAD catalog) and we investigated whether the total intra-night light curve RMS, the white noise, 
and the red noise contributions showed any significant dependence on colour differences between the red M dwarfs and the 
typically bluer sets of comparison stars. In parallel, we investigated the possible dependence of correlated and 
random noise on the average number of comparison stars used for each target, which could also play a role 
to some degree (especially if one or more of the comparison stars were to show variability). 

The left three panels of Fig.~\ref{colordep} show the average red noise (top), white noise (middle), 
and total RMS (bottom) for the sample as a function of the difference between the colour index of the target
and the average color index of the corresponding comparison stars $\textit{(Vr-Kr)-(V-K)}$. The 
data are binned along the X-axis and error-bars correpond to the dispersion of the values in a bin 
(the right-most bin contains only two stars). No evident trends with colour index can be found. 
In the right three panels of Fig.~\ref{colordep} the same quantities are plotted as a function of the
average number of comparison stars for each red dwarf in the sample. Even in this case, there is no evident correlation. 
On the time scale of one observing night, the number of comparison stars or colour-dependent effects such as atmospheric extinction 
do not appear to impact to any significant extent our measurements. This is in line with, e.g., the analysis of 
Bailer-Jones \& Lamm (2003) and Irwin et al. (2011). The same considerations may not hold 
in case of our full-period photometry, which is more prone to being affected by other sources of systematic errors, 
including possible dependency on colour-related effects (e.g., flat-fielding errors), as already mentioned in \S~\ref{phot_prec}. 
An in-depth analysis to search for colour-dependent effects in the full-period light curves will be the objective of future work.

We also looked at the possibility of evaluating the impact on the noise levels by forcing the choice of 
red stars as comparisons. We analyzed the fields of GJ 1167A and GJ 9652A, with 
common proper motion companions at $\sim3^{\prime}$ and $\sim42^{\prime\prime}$, respectively. 
Due to their faintness (at $V$-band they are 2-3 mag fainter than the primaries), GJ 1167B and 
GJ 9652B are not selected as comparison objects by any of the four intra-night differential photometry methods. 
By forcing their inclusion in the reference sets for both targets, no appreciable differences in either 
correlated or random noise levels can be observed, due to the significant number ($>10$) of much brigther references 
selected by the pipeline in all cases. 

While this preliminary analysis has not uncovered significant color-dependent effects which could be 
ascribed to the choice of references, we are aware that prescriptions within TEEPEE should still be included 
for handling comparison sets with average colour indexes not vastly different from that of the target, 
particularly when it comes to reducing data over a long period of observations. Recipes have already 
been identified by other programs targeting specifically large samples of M dwarfs (e.g., Irwin et al. 2011), 
which are fine-tuned to the specific observing strategy, observational setup, and sample size. 
Such a careful data treatment will be applied in the future to our survey data. 

\subsection{Photometric variability: periodicity analysis}

After the characterization of the photometric noise properties for the M dwarfs observed during the pilot study, 
we describe here the key elements of the analysis segment devoted to the astrophysical characterization of the stars 
themselves. We look for signals in the data which are periodic in nature, and which would indicate either the 
presence of a transiting companion or which could be interpreted as intrinsic to the target (e.g., due to chromospheric 
activity). In separate works (Damasso et al. 2011; Giacobbe et al. 2012 in prep.), the search tools applied here to our 
program stars were used to characterize the nature of tens of newly discovered variable stars in the target fields.

\subsubsection{Searching for transit-like events}\label{sec:bls}

While showing an exact periodicity, a highly non-sinusoidal transit event cannot be modelled efficiently using standard approaches 
based on finite sums of sinusoidal components, such as the Discrete Fourier Transform method (e.g., Deeming 1975), 
or on period-finding techniques which minimize the scatter in smoothed light curves, such as the Phase Dispersion Minimization 
algorithm (e.g., Stellingwerf 1978). As is now common practice, in order to detect periodic transit events in our data 
we use a method of least squares fits of step functions to a folded signal corresponding to a grid 
of trial periods, as realized in the Box-fitting Least Squares (BLS) algorithm (Kov\'acs et al. 2002). 

In the analysis (performed in IDL with the exception of a C++ implementation of the BLS period search), we 
utilized (for each photometric reduction method) the intra-night light curves of each target without $\sigma$-clipping, 
as (partial) transit events might in principle be recognized as clusters of outlier points and removed from the datasets.
All light curves were inspected for transit-like signals using a dense grid of 10000 trial periods in the range 0.4-5.0 days. 
We divided the folded time series into 300 bins and we evaluated the signal residue (i.e., the BLS power spectrum) 
of the time series using these binned values. We fixed the fractional transit length in the range of 0.1-0.01. 
The study of the BLS spectra does not show any significant periodicity within the period range investigated, for all objects in our sample. 
This is not unexpected, given the very small number of targets included in the pilot study, due to standard considerations 
of geometric transit probability. 

\subsubsection{Characterizing stellar rotation}\label{sec:rotate}

For most of our program stars, the photometric data collected during our one-year long observing campaign, 
covering a typical timespan of $\sim$2 months, can be used to directly
measure the stellar rotation periods if quasi sinusoidal variations in the broad-band photometric 
signal are detected, under the assumption that it is produced by short- and medium-lived spots on 
the stellar photosphere.  

In order to reveal the presence of presumably starspots-induced rotational modulation in the full-period differential 
photometric datasets (extracted with the $m2$ and $m4$ methods) of our targets, we looked for agreement between 
two different periodicity search algorithms 
(and limiting our analysis to the determination of the most significant frequency). Both algorithms produced 
essentially the same results  when applied on each of the two light curves, making our conclusions more robust. 
In the following discussion we present the results of the period-search analysis as applied to the light-curve 
obtained with the $m4$ method.

The first tool utilized is the PERIOD04\footnote{http://www.univie.ac.at/tops/Period04/} software (Lenz \& Breger 2005), 
which performs Discrete Fourier Transforms (DFT) of a time series and is used 
routinely in asteroseismology work (e.g., Aerts et al. 2010). 
The second method solves directly a linear Least Squares problem: the data are folded according to a grid of 
different trial periods and fit to a sine function, and at each step the reduced chi-square $\chi^{2}$ 
is evaluated. While in the Fourier analysis the significant periods correspond to peaks in 
the amplitude spectrum, in a periodogram obtained with the second method the best-fit period 
will correspond to the trial value which minimizes $\chi^2$. This approach has been adopted by Irwin et al. (2011) 
to search for rotation periods in the M dwarf sample of the MEarth survey.

We found that our program stars can be divided in three main groups: $a)$ targets which do not 
show any significant periodicity, $b)$ targets for which the power spectra show the existence of several 
possible periodic signals, but with low significance, and $c)$ stars showing one significant frequency 
in the periodogram which can be reliably interpreted as the star spinning frequency (2 objects in particular, $\sim$ 10$\%$ of the targets). 

\begin{figure}
\centering
\includegraphics[width=0.47\textwidth]{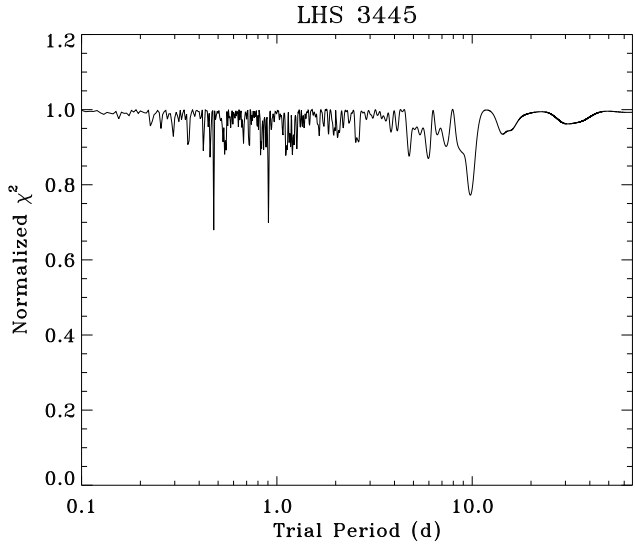}
\caption{The search for periodicities in the full set of the photometric data for the dM star LHS 3445 revealed the existence 
of a signal with period P=0.4752812 days (corresponding to the absolute minimum in this periodogram).}
\label{LHS3445period}
\end{figure}

\begin{figure}
\centering
\includegraphics[width=0.48\textwidth]{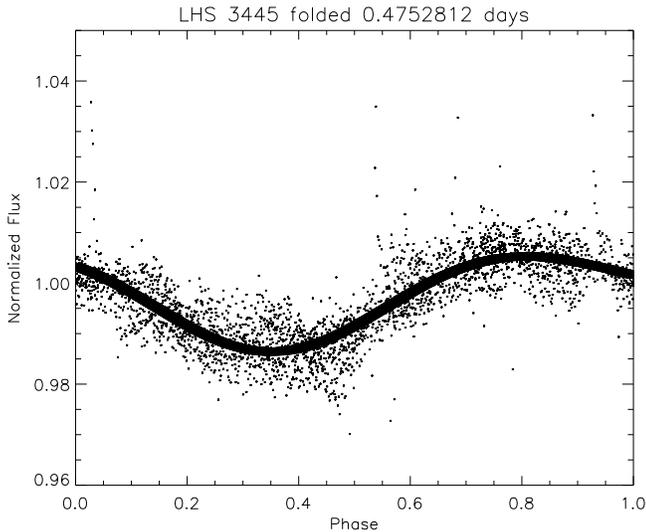}
\caption{Photometric data for the dM star LHS 3445 folded according to its probable rotation period P=0.4752812 days. 
Overplotted is the best-fit single-spot model described in \S~\ref{sec:starspots}}
\label{LHS3445folded}
\end{figure}

Here we present the results concerning the determination of the rotation periods of these two targets, 
LHS 3445 (a.k.a. GJ 9652A) and GJ 1167A. In order to gather independent circumstantial 
evidence for their periodicity being due to rotation, as observed in our photometric data. 
We downloaded archival high-resolution spectra of both targets obtained by Shkolnik et al. (2009) with the High Resolution Echelle 
Spectrograph (HIRES) on the Keck 1 telescope, and we analyzed them to provide 
an estimate of the projected rotational velocity $v\sin i$. In short, the $v\sin i$ values for the stars are measured from the stellar line 
widths via a cross-correlation technique which employs as a template a high-resolution spectrum 
of a slow rotating star of similar spectral type and known projected rotational velocity 
(see, e.g., Reid \& Mahoney 2000). 
The width of the peak in the cross-correlation function is dependent on the line profiles 
of both template and target object. We used a HIRES spectrum of the star GJ 402 
(spectral type dM5) as the slowly rotating template. This spectrum is used to create 
artificially broadened profiles, with an IDL code that implements the prescription of 
Gray (2008), and the width of the cross-correlation function (CCF) peak between the unbroadened template and its broadened 
version is calibrated for a set of rotation speeds. 
The maximum in a CCF peak vs. $v\sin i$ plot will correspond to 
the best-fit projected rotational velocity for the target. For our analysis 
we chose the echelle order covering the spectral range 7370-7490 \AA, 
which is free of TiO bands and telluric lines which would make the analysis difficult.

\begin{figure}
\centering
\includegraphics[width=0.47\textwidth]{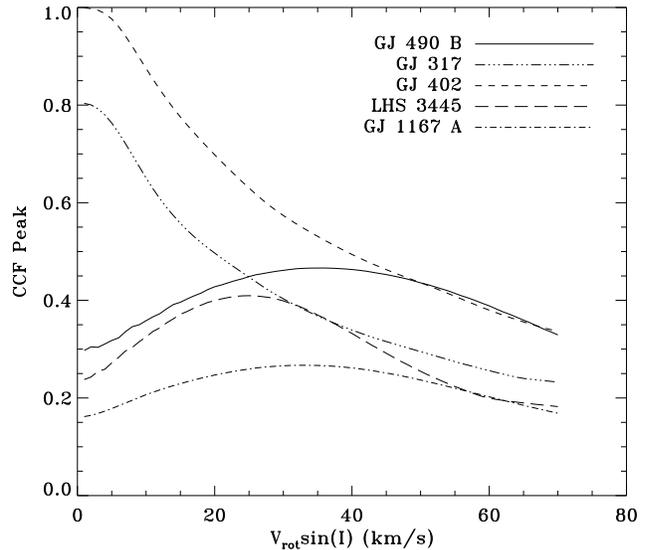}
\caption{Cross-correlation function (CCF) peaks as a function of the projected rotational velocity $\textit{v}$sin($\textit{i}$) 
for two of our dM targets, LHS 3445 and GJ 1167A, and other M dwarfs used as a template (GJ 402) and check stars used to 
validate the procedure. The maximum values of the CCF peak functions correspond to the projected rotation velocities 
measured from the spectra taken with the Keck/HIRES echelle spectrograph. LHS 3445 and GJ 1167A appear to be rapid rotators.}
\label{CCFvsini}
\end{figure}

The star LHS 3445 is included in the samples analyzed by Shkolnik et al. (2009) and Walkowicz \& Hawley (2009), 
and it is known to be an X-ray active star (it has been identified in ROSAT data) and a flare star 
(see table\ref{table:3}). We observed this target flaring three times, as described in \S~\ref{sec:flare}. 
LHS 3445 is also considered a young star: Smart et al. (2010) indicate an estimated age in the range 30-50 Myr, 
while Shkolnik et al. (2009) suggest an age between 60 and 300 Myr. Fig.\ref{LHS3445period} displays the 
calculated periodogram for LHS 3445, 
showing a minimum occurring at $\sim$0.47 days, and this is confirmed by the DFT analysis. 
This is then to be considered the most probable rotation period.
A second minimum is observed very close to the period of 1 day and we discard it as an alias on the basis 
of the window function. Fig.\ref{LHS3445folded} shows our photometric data folded according to the period P=0.4752812 days, 
revealing a clear sinusoidal-like shape probably due to star spot modulation. A single-spot model (see \S~\ref{sec:starspots}) is overplotted 
to highlight the consistency of this hypothesis. 
As shown in Fig. \ref{CCFvsini}, the analysis of the Keck spectrum leads to the best estimation for the projected rotational velocity 
$v\sin i\sim$25 km/s. Note in Fig.\ref{CCFvsini} that the slow rotating star GJ 317 
($v\sin i < 2.5$ km/s; Browning et al. 2010) is used as a 
check for the goodness of the results for our targets (i.e. we were able to recover its low projected 
velocity analyzing the HIRES spectrum) together with the star GJ 490 B, known to be a fast rotator 
($v\sin i = 41$ km/s; Phan-Bao et al. 2009), for which we recover a high value of $v\sin i$ in agreement with the one 
found in the literature. Using the stellar radius indicated in Table \ref{table:3} and the rotation period estimated photometrically, 
the mean rotational velocity of LHS 3445 results to be 44.7 km/s, which compared to the projected 
velocity leads us to conclude that the rotation axis is inclined by $\sim$34 deg with respect to the line of sight.   
\footnote{Note that LHS 3445 has an M dwarf common proper motion companion, the star LHS 3446. 
We analysed its light curve, which appears to be not at all variable over the timescale of our observations.}

\begin{figure}
\centering
\includegraphics[width=0.47\textwidth]{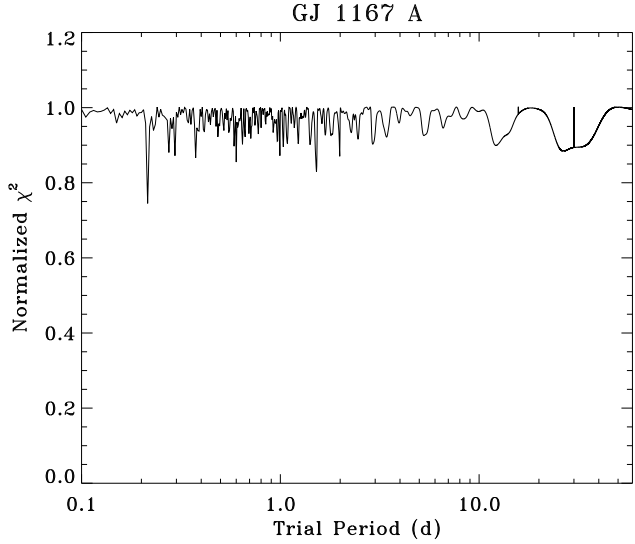}
\caption{The search for periodicities in the full set of the photometric data for the dM star GJ 1167A revealed the 
existence of a signal with period P=0.2151997 days (corresponding to the absolute minimum in this periodogram).}
\label{GJ1167period}
\end{figure}

\begin{figure}
\centering
\includegraphics[width=0.48\textwidth]{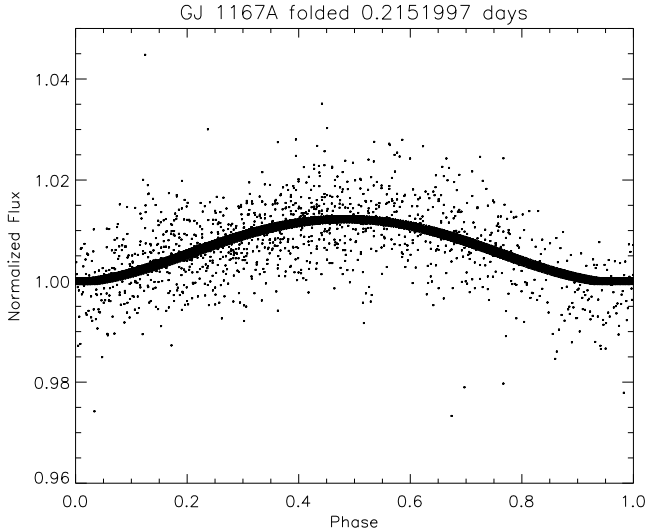}
\caption{Photometric data for the dM star GJ 1167A folded according to its probable rotation period P=0.2151997 days. 
Overplotted is the best-fit single-spot model described in \S~\ref{sec:starspots}}
\label{GJ1167folded}
\end{figure}

The second target that shows a clear low-amplitude sinusoidal modulations in the light curve is GJ 1167A.
GJ 1167A is an active flare star (but no flares were observed over the timespan of our observations), 
included in the samples analysed by Shkolnik et al. (2009), who classify 
it as a young dM4.8 star with an age in the range 60$\div$300 Myr. In Smart et al. (2010) GJ 1167A is reported being 
a dM4.0 star with a rather uncertain (0.12$-$10 Gyr) age. Both the DFT and PDM analysis of 
our photometric data reveals a clear sinusoidal periodic signal corresponding to P$\sim$0.22 days (the PDM results 
are shown in Fig. \ref{GJ1167period}). Fig. \ref{GJ1167folded} shows the light curve folded according to this period, with superposed a 
single-spot model as for LHS 3445 (see \S~\ref{sec:starspots}). 
The fact that GJ 1167A is a very fast rotator is confirmed by the line broadening analysis of the 
HIRES spectrum, as shown in Fig. \ref{CCFvsini}. We estimate $v\sin i\sim 33$ km/s. 
Assuming for this star a radius $R=0.14R_{\odot}$, our measured photometric rotation period implies a 
true rotational velocity of 32.9 km/s, which means we are looking at GJ 1167A virtually perpendicularly to its rotation axis.

Given the high rotational velocities inferred for both LHS 3445 and GJ 1167A, this possibly implies relatively young ages 
for these stars. Alternatively, the rapid spin period could also be due to tidal synchronization effects if these stars 
were to be very short-period binaries. Indeed, Smart et al. (2010) hint at the possible binarity of LHS 3445 based on 
kinematics considerations. We checked the cross-correlation functions obtained from the archival high-resolution Keck/HIRES 
spectra, looking for evidence of double peaks, which would indicate the presence of the secondary spectrum. 
No such evidence is found in both cases, thus we tentatively conclude that LHS 3445 and GJ 1167A do not harbor 
very short-period companions of comparable mass. Naturally, a more aggressive analysis could be carried out, 
based for example on detailed modeling using improved cross-correlation techniques such as TODCOR (Zucker \& Mazeh 1994), 
aimed at faint companions detection. We plan to pursue this issue further in future work. 

\subsection{Photometric variability: starspots analysis}\label{sec:starspots}

On the basis of the results described in the previous paragraph, we attempted at modeling the observed 
photometric variability of LHS 3445 and GJ 1167A assuming it is due to a rotating spot on the stellar photosphere. 
This work must be intended as the first step towards a more detailed model for star spot 
distribution to be developed in the course of the upcoming long-term photometric monitoring program, 
which will become a particularly valuable tool to evaluate the impact on the low-mass planet detection thresholds (e.g. Barnes et al. 2011).

The model we developed is based on the work of Makarov et al. (2009). This three-dimensional model describes the 
flux modulation due to a small circular spot rotating on the stellar surface. We modelled a single 
circular spot of radius \emph{r} with its center located at latitude \emph{b} and longitude \emph{l} 
in the frame of reference of the star. The star is assumed rotating around its axis with a differential 
angular velocity $\vec{ \omega}(b)$ which depends on the latitude on the stellar disk, and the spot is 
assumed rotating at a fixed latitude. The rotation axis is considered tilted by an angle \textit{i} measured 
starting from the line of sight ($i=0$ if the axis is pointing toward the observer; $i=\pi/2$ when the axis 
is perpendicular to the line of sight). In the model, the flux modulation can be expressed as 
$\frac{\Delta F}{F}=-f_s \frac{r^2I(\theta)\cos \theta}{I_{tot}}$, where $f_s$ is the spot 
contrast with respect to the local surface brightness, \textit{r} is the spot size in units of the stellar radius ($r\ll1$), 
$I_{tot}$ is the integrated intensity flux from the stellar disk, $\mathbf{\theta= \theta (b,l(t))}$ is the angle 
between the line of sight and the perpendicular to the star surface passing through the center of the 
spot (depending upon the epoch of observation), and I($\theta$) is the intensity flux emitted at the 
spot location. Using the Stefan-Boltzmann law $I=\sigma T^4$, the contrast can be approximated by the 
relation $f_s=1-\frac{T^4_s}{T^4_{ph}}$, where $T_{s}$ and $T_{ph}$ are the mean temperatures of the 
spot and photosphere respectively. Using literature prescriptions (Barnes et al. 2011), we set the contrast to vary in the range 0.2-0.8.

The parameters kept fixed in the simulation are the $I$-band limb-darkening coefficients (a quadratic limb-darkening 
law from Claret (2000) appropriate for M dwarf stars is used to describe the intensity flux distribution on the stellar photosphere) 
and the orbital period. The free parameters are the inclination angle of the rotation axis, the spot latitude and the initial 
longitude, and the scalar $f_s\cdot$r$^2$ that, combining the spot contrast with its surface area, quantifies 
the dimming in the light intensity due to the transit of the star spot in front of the observer.

When applied to the dM stars LHS 3445 and GJ 1167A, the model reproduces well in both cases the 
inclination of the rotation axis, as found from spectro-photometric data: $i_\star=26.7$ deg for LHS 3445, 
and $i_\star=89.9$ deg for GJ 1167A. The post-fit residuals, having RMS of 4.2 mmag and 6.4 mmag for 
LHS 3445 and GJ 1167A, respectively, are comparable to the typical RMS precision achieved 
for the sample (see \S~\ref{phot_prec}), and do not show any additional significant evidence for periodic signals. 
All best-fit spot model parameters for the two stars are summarized in Table~\ref{table:7}, and the 
corresponding flux modulations are overplotted to the phased photometric data for the two stars in Fig.~\ref{LHS3445folded} and Fig.~\ref{GJ1167folded}.

\begin{table}
\caption{Results of the star spot model applied to the fast rotators
LHS 3445 and GJ 1167A}             
\label{table:7}      
\centering                          
\begin{tabular}{c c c c}        
\hline\hline                 
Star&$i_\star$ & $b$ &$f_s\cdot$r$^2$\\
&(deg)&(deg)&\\
\hline
LHS 3445&26.7&20.3&0.00027\\
GJ 1167A&89.9&42.1&0.000007\\
\hline\hline
\end{tabular}
\end{table}

We note that our choice to fit for the scalar $f_s\cdot$r$^2$ rather than 
the individual values for the brightness contrast $f_s$ and the spot
radius $\textit{r}$ is due to the fact that in the Makarov et al. (2009) model these 
two parameters appear strongly correlated, and attempting to fit for them 
separately results in very loose constraints on their actual values, and 
poor overall convergence of the spot model. 
This effect can be easily understood, as in the disk-integrated stellar flux 
data the effect of a cold, small spot is difficult to distinguish from the signal 
produced by one which is warmer and larger. The price to pay in this case, 
in that we don't directly estimate the spot size and temperature, is minimal, 
as this exercise, for the purpose of this study, must be intended solely
as a consistency check of the interpretation of the rotational modulation 
for the two stars based on the photometric and spectroscopic data presented above. 
Possible future developments include the implementation of more sophisticated (single and multiple) 
spot models, and the detailed assessment of their uncertainties through, e.g., Bayesian statistics.

\subsection{Photometric variability: flares analysis}\label{sec:flare}

Flaring events, a powerful indicator of the presence of strong stellar 
magnetic fields, are known to be relatively common among late-type stars, 
typically lasting between a few minutes up to several hours, and producing increases 
in the observed flux of up to several magnitudes. 
While ground-based as well as space-borne long time-series photometric studies 
of open clusters and in the field over a range of wavelengths 
(Moffett 1974; Lacy et al. 1976; Audard et al. 2000; G\"udel et al. 2003) have 
begun unveiling some important correlations between flare occurrence 
rates and stellar characteristics such as mass, age, and activity levels 
(e.g., Ambartsumyan et al. 1970; Mirzoyan et al. 1989; Kowalski et al. 2009), 
the physics of flares is still far from being fully understood. Several 
open questions still need to be properly addressed, which include the details
of the energy release, the mechanisms for producing the
atmospheric emission, and the understanding of flares on a global
scale -- how do flare properties (occurrence rates, emission strength, timescales, frequency) 
correlate with stellar characteristics (mass, age, activity levels)?

Upcoming ground-based photometric surveys such as Pan-STARRS 
(Kaiser 2004), PTF (Law et al. 2009), and LSST~\citep{abell09} will certainly provide 
the opportunity to gather flare data for large numbers ($>10^6$) of stars, sampling 
wide ranges of flare amplitudes and timescales. However, a potentially important 
niche for `classical' flare studies (i.e., those based on the continuous monitoring 
of single objects) will come as a by-product of those photometric programs, such as MEarth 
(Nutzman \& Charbonneau 2008) or the upcoming survey at OAVdA, 
targeting large numbers of relatively bright late K and M dwarfs in search of transiting planets. 
For example, the detailed characterization of the flaring behavior for K and M dwarfs 
objectives of targeted searches for transiting planets is an important 
ingredient towards the thorough understanding of the impact a star and its environment 
might have through time on the habitability of any planet it may harbor which could sustain 
the presence of liquid water on its surface (e.g., Kasting et al. 1993; Lammer et al. 2007; 
Guinan et al. 2009). 

\begin{figure}
\centering
\includegraphics[width=0.45\textwidth]{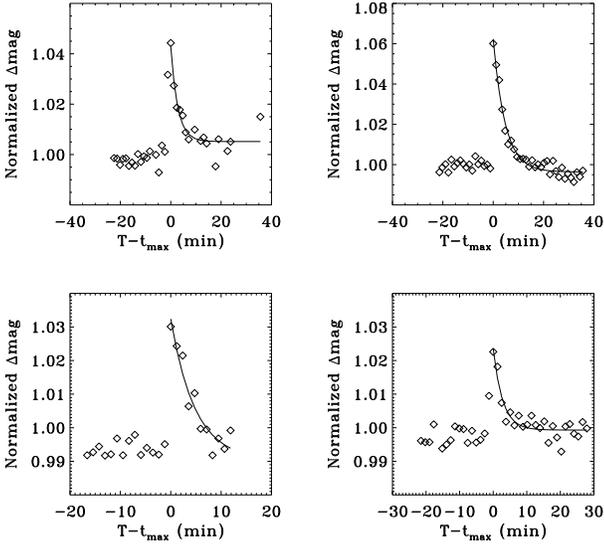}
\caption{Four flares seen in the LHS 3445 (upper and lower left panels) and LHS 2686 (lower right panel) light curves. 
In each case, the solid line shows the fit of Equation~\ref{eq:flare} to the light curve. For LHS 3445, 
two almost consecutives flares were recorded, with approximately equal decay times of $\sim4.5$ min.}
\label{figflares}
\end{figure}

Two of the stars included in our pilot study, LHS 3445 and LHS 2686, showed 
flaring events in the photometry gathered over a period of 66 days 
and 60 days, respectively. LHS 3445 is classified as UV Cet type flare star (Gershberg et al. 1999), 
while no information on flaring is available in the literature for LHS 2686. Given the sampling 
rate of the two light-curves ( 54 sec and 75 sec, respectively), we were sensitive to 
flares with decay times larger than $\sim1$ min. Visual inspection of the light-curves allowed us to 
infer decay times of a few minutes (`impulsive' flares. See, e.g., Krautter 1996). 
As an illustrative exercise of the type of studies that will be possible once observations will 
be gathered in survey mode, we characterized the events following the 
approach of Hartman et al. 2011, i.e. by solving a non-linear Least Squares problem with 
a set of consecutive photometric measurements starting from the time of recorded maximum brightness 
$t_0$ and an exponential model function of the type: 

\begin{equation}
m_d(t) = A\times\exp{-(t-t_0)/\tau}+m_{d,0},
\label{eq:flare}
\end{equation}

with adjustable parameters $A$, the peak magnitude of the flare relative to the non-flaring 
magnitude, $\tau$, the decay timescale, and $m_{d,0}$, the differential normalized magnitude (see \S~\ref{sec:metod} for 
its definition) of the star before the flare. 
The Least Squares solution was obtained with an IDL implementation of the Levenberg-Marquardt method 
(Levenberg 1944; Marquardt 1963), {\tt MPFIT} by Craig Markwardt\footnote{Available at http://purl.com/net/mpfit. 
{\tt MPFIT} is a port of MINIPACK-1 from FORTRAN, and is also available in C and Python} (Markwardt 2009), 
and starting guesses for the model parameters $A$ equal to the peak magnitude, $\tau=0.001$ d, amd $m_{d,0}$ 
equal to the average differential magnitude of each time series. The results are shown in Figure~\ref{figflares}. 
For LHS 3445, two of the three recorded flares occurred within a timespan of 1.2 hrs during the same night, 
possibly a case of homologous flares (e.g., Martres et al. 1984; Doyle et al. 1990; Ranns et al. 2000). 

\subsection{Limits to transiting companions}\label{limits}

While no transit event has been recorded in our dataset, as reported in \S~\ref{sec:bls}, the photometric 
measurements obtained for our sample can be utilized to carry out simulations aimed at determining 
what sensitivity to transiting companions (of given radius and period) we achieved on a star-by-star basis, 
expressed in terms of easy-to-interpret comparison metrics, such as detection probabilities and phase coverage. These simulation tools. 
whose application to the data collected during the pilot study we illustrate here, will be of great use in 
the careful prioritization and for the optimization of the scheduling of the targets during survey operations. 

In order to evaluate the sensitivity of each of the M dwarfs in our sample to transiting companions 
of given period and given radius, provided the primary size from Table~\ref{table:3} (or equivalently inducing 
a transit signal of given depth), a large-scale simulation was performed by injecting synthetic transit signals 
into the differential light curves (both intra-night and full-period) of our targets (without $\sigma$-clipping as in 
\S~\ref{sec:bls}).
We utilized these data under the hypothesis that the procedure adopted to extract it 
from the raw images does not affect the signal. This simplificatory assumption is often adopted when dealing with 
such issues (e.g., Irwin et al. 2011). Furthermore, transit signal injection was performed on all light curves 
obtained with the four methods for differential photometry described in \S~2. We are aware that the specific choices 
for these elements of the simulation setup can in principle impact our findings to some extent. For example, 
we are not in a position to estimate quantitatively the effects on our capability to recover transit signals 
due to the application of trend-filtering algorithms during the production 
of the differential photometric data in pipeline mode. However, the goal of the present analysis is simply to gauge 
our ability to detect transit events as a function of the system parameters, given an observed distribution of RMS 
values for our stellar sample. The present simulation setup is thus appropriate for this purpose, and any in-depth 
evaluation of, e.g., the quality of the calibration procedures against transit signal detection thresholds 
goes beyond the scope of this paper. 
We plan to relax the above assumptions and caveats in a future work in which we will carry out new studies based 
on a fully consistent end-to-end analysis (including extraction and differential photometry) of modified images in 
which synthetic transit signals will have been injected already at the pixel level. 

The input parameters (and their ranges) to the simulation, which was entirely carried out in IDL, 
we determined based on considerations taking into account the temporal sampling, total timespan, and typical 
photometric precision of our data. We generated:
\begin{itemize}
\item 1000 random, uniformly distributed periods in the range $0.5-5$ days;
\item 100 random, uniformly distributed phases for each period;
\item 4 amplitudes of the transit signal depth $t_d$ (in flux units: 0.02, 0.015, 0.01, 0.005) for each period and phase. 
The four values of transit depth simulated correspond to companion radii in the ranges 1.5-9.1, 1.3-7.9, 1.1-6.4, and 0.8-4.5
$R_\oplus$ , respectively, given the sizes of the primaries from Table~\ref{table:3}; 
\item a fixed orbital inclination $i=90^\circ$, and perfectly circular orbits ($e=0.0$);
\end{itemize}

Consequently, the orbital radius parameter was derived from $M_{*}$ (see \S~\ref{sec:stellar}) and from Kepler's third law 
under the assumption $M_{p} \ll M_{*}$ (Seager \& Mallen-Ornelas 2003). We recognize that allowing the inclination to float 
might change to some extent the details of the detection probability results. A detailed study of the sensitivity to 
non-central (eventually in the limit of quasi-grazing) transits is left for the future.

With the above simulation setup, 400\,000 synthetic transit light curves were generated for each target 
using the Mandel \& Agol (2002) algorithm. For the purpose of this analysis limb-darkening 
effects are essentially irrelevant and in order to speed up execution they were turned off. For each target, 
the synthetic transit signals were then injected on both nightly and full-period light curves.  
The choice of running the simulation for the whole suite of differential photometry methods was driven by the findings 
of \S~\ref{phot_prec} and \S~\ref{rdn}, thus allowing us to gauge the sensitivity of the transit detection probability 
to variable photometric precision. This was done in practice by comparing the results obtained by running the BLS 
algorithm on the synthetic, `noise free' datasets as well as on the combined light curves, with the prescription of 
discarding datasets in which a transit would have occurred only once. 

\begin{figure}
\centering
\includegraphics[width=0.47\textwidth]{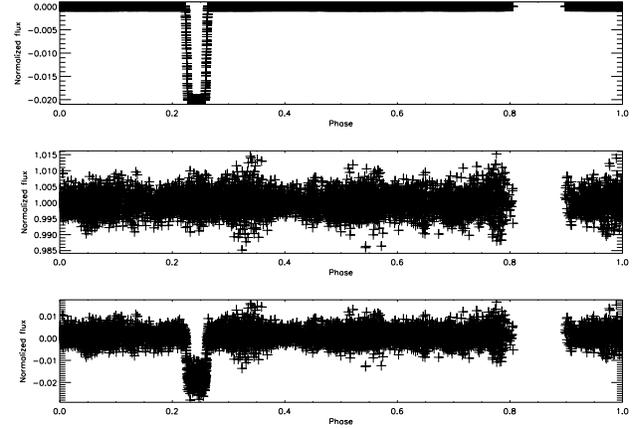}
\caption{Top: synthetic folded light curve with $t_d=2\%$ and a period of 1.8468372 days. The times of observation are 
those of LHS 1976. Center: the actual phased light curve of LHS 1976. Bottom: the combined light curve for LHS 1976} 
\label{figtra}
\end{figure}

Figure~\ref{figtra} summarizes the process of injection of a transit signal in a differential light curve. 
The top panel shows the synthetic folded light curve, without limb-darkening, produced by the Mandel \& Agol (2002) formalism for 
a transit with a fractional depth of 2\% and a period of 1.8468372 days. 
The center panel shows the actual phased light curve of LHS 1976, while the bottom panel shows 
the combined light curve in phase. The datasets shown in the upper and lower panels are both fed to BLS.

In the analysis with the transit search algorithm of a given light curve, BLS was run using 500 period steps in the range $0.5-5$ days, 
the folded time series was divided into 100 bins, and the signal residue ($SR$) of the time series (see Kov\'acs et al. 2002) 
was determined using these binned values. We fixed the fractional transit length to lie in the range $0.1-0.01$.

As mentioned above, the main results of the simulation are expressed in terms of two simple comparison metrics, 
which we identified to be the phase coverage and the transit detection probability.  

Here we calculate the phase coverage, in a period range of 0.5 to 5 days, 
for 10000 trial periods. We divide the phased light curve in bins corresponding to 20 min in length (obviously 
the number of bins changes with the trial period). A single bin is flagged as 'filled' (i.e. containing a satisfying number of data points) if at least five points 
coming from three different nights fall within. The phase coverage (expressed as percentile) represents the relative number 
of bins that satisfy this condition with respect to the total number of bins. 

We define the detection probability as the relative number of periods that are detected by BLS with respect to the total number of injected periods.
We considered as detections all those signals for which BLS returned a period $T^\prime$ such that 
$|T^\prime-T_\mathrm{in}|/T_\mathrm{in}<0.01$, where $T_\mathrm{in}$ was either the actual period of 
the injected signal, or half that period, or twice that period. We are aware of the fact that this 
requirement might be somewhat stringent: In practice we do not consider as detections transits identified 
with the correct depth, but with an incorrect period due to, e.g., the sampling properties of the light curve. 
We adopted this more conservative approach to the definition of detection probabilities as a way to partly 
compensate for the caveats and assumptions of the simulation scenario listed at the beginning of this section.

\begin{figure}
\centering
\includegraphics[width=0.5\textwidth]{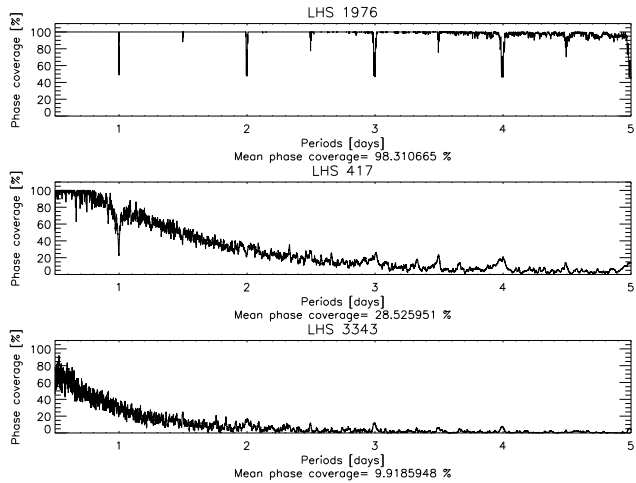}
\caption{Phase coverage as a function of period of the injected transit signal ($t_d=2\%$). Top: LHS 1976. Center: LHS 417. Bottom: LHS 3343} 
\label{figph}     
\end{figure}

Figure~\ref{figph} shows the phase coverage as a function of injected period for a signal with $t_d=0.02$ 
for three representative objects in our sample, LHS 1976, LHS 417, and LHS 3343, with very different 
(from poor to excellent) levels of phase coverage.

\begin{figure*}
\centering
\includegraphics[scale = 0.3, angle=-90]{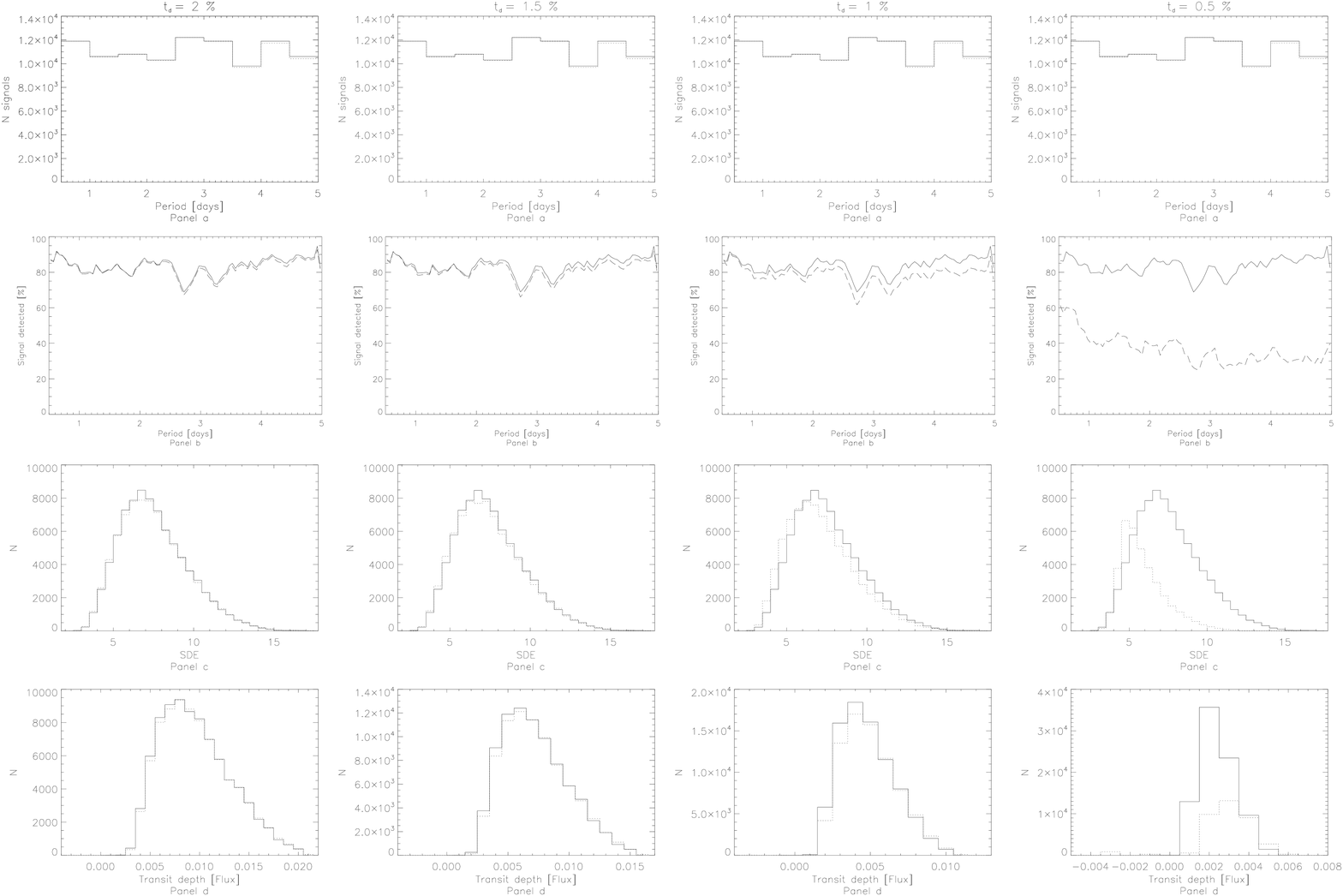}
\caption{Summary of simulation output for LHS 1976 (see text for details). 
Panels $a$: number of signals vs period; Panels $b$: detection probability vs period;
Panels $c$: SDE distributions; Panels $d$: distributions of transit depths of the detected signals.} 
\label{figan1}     
\end{figure*}

\begin{figure*}
\centering
\includegraphics[scale = 0.3, angle=-90]{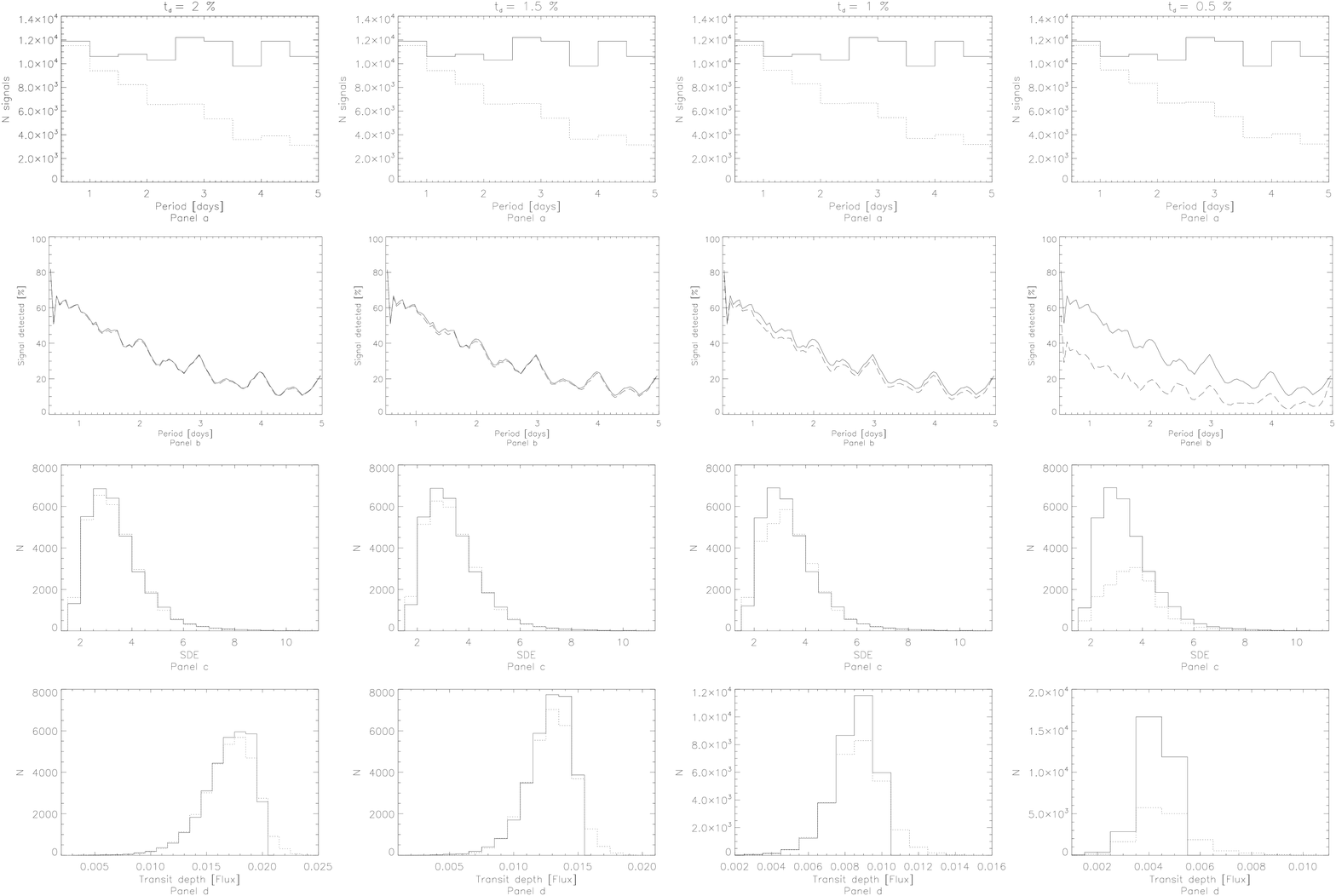}
\caption{Summary of simulation output for LHS 417 (see text for details). 
Panels $a$: number of signals vs period; Panels $b$: detection probability vs period;
Panels $c$: SDE distributions; Panels $d$: distributions of transit depths of the detected signals.} 
\label{figan2}     
\end{figure*}

\begin{figure*}
\centering
\includegraphics[scale = 0.3, angle=-90]{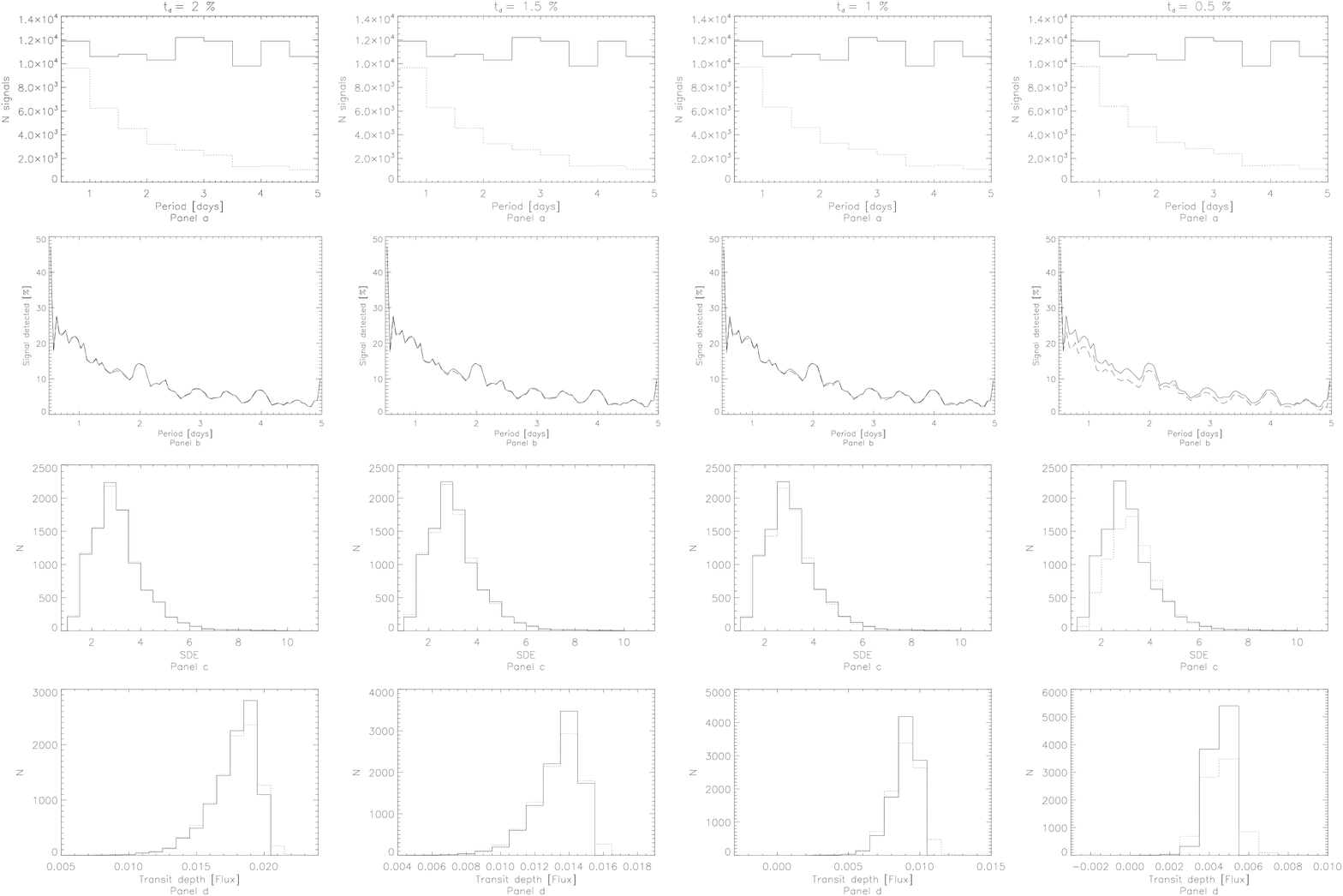}
\caption{Summary of simulation output for LHS 3343 (see text for details). 
Panels $a$: number of signals vs period; Panels $b$: detection probability vs period;
Panels $c$: SDE distributions; Panels $d$: distributions of transit depths of the detected signals.} 
\label{figan3}     
\end{figure*}

For the same three stars, and in the case of differential light curves obtained with method $m3$, 
Figure~\ref{figan1}, Figure~\ref{figan2}, and Figure~\ref{figan3} give an overview 
of all the fundamental ingredients provided in output from the simulation. In every Figure, the top-most four 
sub-panels ($a$) show, for each of the values of $t_d$ simulated, the original distribution of injected transit signals 
(solid line) and that of those signals whose corresponding light curves (synthetic as well as combined) 
are subsequently fed to BLS (dashed line), both expressed as a function of period. 
The second set of sub-panels ($b$) shows the fraction of BLS-detected signals in the synthetic datasets (solid line) and 
in the combined light curves (dashed line). The third set of sub-panels ($c$) shows the distribution of the 
signal detection efficiency ($SDE$) parameter both in the case of synthetic light curves (solid line) and for the 
combined light curves (dashed line). We remind the reader that a value for $SDE$, which quantifies the 
statistical robustness of a detected transit-like periodic signal in the language of BLS, is computed as:

\begin{equation}
SDE = \frac{SR_{peak}-\langle SR \rangle}{sd(SR)},
\label{eq:sde}
\end{equation}

where $SR_{peak}$ is the value of the peak in the signal residue distribution, while $\langle SR \rangle$ and $sd(SR)$ are 
the mean and the standard deviation of \textit{SR} over the frequency band tested (see Kov\'acs et al. 2002 for details). 
Finally, the last set of sub-panels ($d$) shows the distribution of the transit depths of the detected signals 
as evaluated by BLS from the synthetic light curves (solid line) and from the combined light curves (dashed line).

For all three Figures, sub-panels \textit{a} and \textit{b} clearly illustrate the dependence 
of transit detection probability on period. As expected, this behaviour is very similar to the 
trend of phase coverage with period (see Figure~\ref{figph}). Panels $b$ also quantify the amount of 
degradation in detection probability, with respect to the 'noise-free' case, when the transit 
depth approaches the typical photometric precision, even when the phase coverage is very good.

The three sets of sub-panels \textit{c} illustrate how the statistical significance of a detection, 
as measured by the $SDE$ values, changes with phase coverage and signal depth. 
The values of $SDE$ can also be used to quantify the probability of a detection being a false 
positive. The actual behaviour of $SDE$ depends on many factors, some intrinsic to the algorithm 
(e.g. number of bins, number of trial periods) and others depending on the quality 
of the data and of the signal (e.g. $S/N$), however the statistical studies Kov\'acs et al. (2002) 
indicate a value of $SDE\sim5$ above which the false positive probability is low ($< 10$\%). 
Above such value the $SDE$ does not depend on the input parameters of BLS (Figure~5 of Kov\'acs et al. 2002). 
The plots in Figure~\ref{figan1} show how the typical value of $SDE$ decreases with the amplitude of the signal and with 
the phase coverage. In Figures~\ref{figan2} and~\ref{figan3} this trend is less evident. 
This is due to the $SDE$ values fluctuating within a regime where the false positive probability is very high. 
The fact that many signals are detected with a low $SDE$ highlights the robustness of the BLS 
algorithm, which is capable of uncovering weak and noisy signals albeit with low statistical confidence.

The analysis of the transit depth distribution, as shown in panel \textit{d}, is 
a further test of the BLS reliability. Generally BLS provides a good estimate of the signal amplitude, but 
looking at the three figures an underestimation of the signal amplitude can be noticed. This is due 
to the initial parameters (number of bins, number of trial periods) of BLS that, in this case (in order to reduce the CPU time calculation), are 
insufficient to define precisely the `border' of the signal. This effect appears to increase with the phase 
coverage. This should be explained taking in account that the number of "spurious" points increases with such phase coverage. 
To verify that this underestimate depends on the input parameters of BLS, we ran the algorithm on our archive data
for the transiting planet WASP-3b (Damasso et al. 2010) increasing the number of trial periods to 10000 (in the same range of the simulation) and 
the number of bins to 300, and we recovered a transit depth in good agreement with the published one (nearly 2\%).

\begin{figure}
\centering
\includegraphics[width=0.5\textwidth]{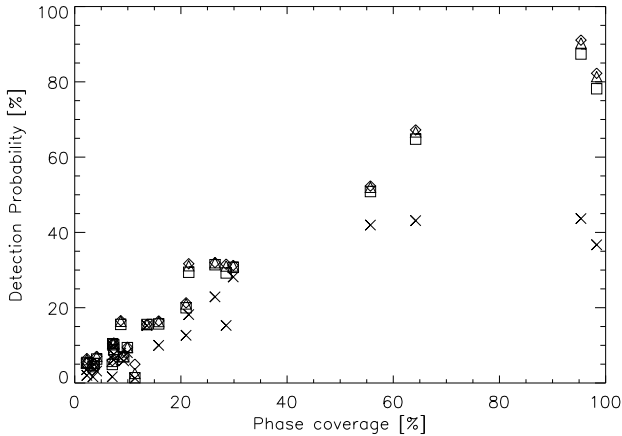}
\caption{Average detection probability as a function of average phase coverage over the 
period range 0.5-5.0 days. Different symbols are used to show the trend of detection probability 
for the four regimes of transit depth simulated ($t_d=2.0$\%: diamonds; $t_d=1.5$\%: triangles; 
$t_d=1.0$\%: squares; $t_d=0.5$\%: crosses).} 
\label{figfas}     
\end{figure}

The previous considerations allow us to relate directly the two comparison metrics 
whose properties we have analysed here, i.e. transit detection probability and phase coverage. 
We show in Figure~\ref{figfas}, for the full stellar sample under investigation and using the simulation 
results based on light curves obtained with method $m3$, the detection probability as a 
function of the phase coverage, both averaged over the whole period range $0.5-5.0$ days. 
Different symbols are used to show the trend of detection probability for the four regimes of 
transit depth simulated. For the definitions of transit detection probability and phase coverage 
provided above, the plot suggests a limit to the probability of detecting a transit, at least utilizing 
a real-life detection algorithm such as BLS, even when the phase coverage 
is almost 100\%, a limit which becomes severe when the magnitude of the signal approaches the typical 
precision of the photometry. 
We believe that this effect is due to several factors, which include (but are not limited to): 
a) the way we actually define a detection: As discussed above, we adopt here a conservative definition, 
likely contributing to missing some of the correctly identified transit depths simply because the period did not fall 
within the stringent agreement constraints required; b) the fact that detection probabilities are averaged over period 
ranges which include values (close to integer and half-integer days) not optimally sampled in a ground-based, single-site 
campaign (see Figure~\ref{figph}); and c) the way we actually define the phase coverage: The choice of bin size and 
the prescription for the number of points in a bin required to define a specific phase covered can also impact the 
likelihood that the transit will be detected at that specific phase.

\begin{figure}
\centering
\includegraphics[width=0.5\textwidth]{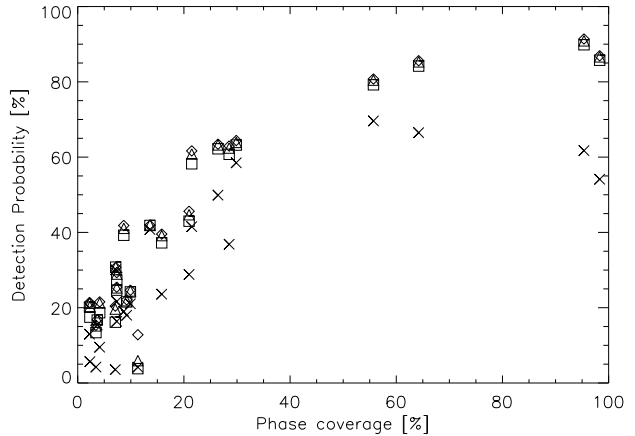}
\caption{Average detection probability as a function of average phase coverage for $P<1.0$ days. 
All symbols are as described in Figure~\ref{figfas}.} 
\label{figfas1}     
\end{figure}

Similarly, Figure~\ref{figfas1} shows the trend of detection probability as a function of phase 
coverage when limiting ourselves to the period range 0.5-1.0 days. Overall, for objects with 
good ($>50\%$) phase coverage, we would have had $>80\%$ chances of detecting transiting companions 
with transit depths in the range $0.5\%<t_d\leq2\%$. Given the estimated stellar radii for these stars, 
this translates in a sensitivity to companions with minimum radii in the range $\sim1.0-2.2$ $R_\oplus$. Figure~\ref{figfas1} 
also shows how, when enough transits ($\gg3$) are observed, even signals of magnitude comparable to 
the photometric precision can be reliably retrieved in our data. 

As mentioned earlier, the results shown here cover the analysis of the simulation based on light curves 
derived with the $m3$ photometric analysis method. 
The results based on the other three intra-night methods (not shown here) are very similar 
while they worsen significantly when the full-period light curves are considered.
This is an expected result because those methods produce more noisy light curves (see \S~\ref{sec:photprec}).
However, such an effect is cause of no significant worries, as operationally transit-like events, whose 
duration for objects in short periods does not exceed the typical length of observations during one single night, are 
to be searched for in nightly-reduced differential photometric datasets. 

To conclude, two points are worth mentioning further here. First, in a large-scale survey of 
thousands of M dwarfs, while some degree of prioritization of the targets based on their 
measured and/or inferred variability properties (based, e.g., on activity indicators and rotation 
information) will be possible, for many stars disregarding variability will simply not be an option. 
As for the two stars for which we successfully determined rotation periods, the probability of 
detection of a transit is not affected by the presence of the rotational modulation effect, as in both 
cases the time-scales of central transit events probed in our simulations (between 0.8 hr and 1.8 hr for LHS 3445 and 
between 0.4 hr and 0.9 hr for GJ 1167A) are significantly shorter than the 
observed rotational periods. Certainly, for much longer transit durations, or much shorter rotational 
periods, variability can instead become a matter of concern as it would directly interfere with a transit 
search. Rotational modulation on time-scales of 0.1 days is indeed observed (e.g., Irwin et al. 2011), and such issue 
will be addressed and dealt with in detail in the future. 
Second, variable degrees of correlated noise, either instrumental (tracking, flat-fielding), 
environmental (airmass, absorption, seeing) or astrophysical (colour, variability) in nature, can also affect 
our capability to detect transits. As shown by Pont et al. (2006), red noise primarily impacts the 
significance thresholds with which a transit event can be recovered from the data in a survey. 
As discussed in \S~\ref{rdn}, it constitutes a challenge to be able to fully characterize the 
relative role played by correlated noise sources on the potential of a photometric survey to 
detect statistically significant transit events, as this analysis will also depend on the details of 
the observed target populations. In the future, such an assessment will be made using our survey data and 
a statistically significant sample of targets.

\section{Summary and conclusions}

We report results of a one-year long photometric monitoring campaign of a
sample of 23 nearby ($d < 60$ pc), bright ($J < 12$) M dwarfs carried out in Italy at the Astronomical
Observatory of the Autonomous Region of the Aosta Valley (OAVdA), 
using small-size ($<1$-m class) telescopes. This survey was conceived as a necessary
preparatory step towards a long-term search for transiting, small-radius 
planets around thousands of dM stars, which will be conducted at OAVdA 
with an array of automated 40-cm telescopes, beginning early 2012. 
This `pilot study' was designed to achieve two goals: $1)$ demonstrate the sensitivity
to $<4$ $R_\oplus$ transiting planets with periods of a few days around our program stars, through a
two-fold approach that combines a characterization of the statistical noise properties of
our photometry with the determination of transit detection probabilities via simulations,
and b) where possible, improve our knowledge of some astrophysical properties (e.g.,
activity, rotation) of our targets through a combination of spectroscopic information and our
differential photometric measurements. At a technical level, the results we obtained 
during the pilot study are instrumental to the accurate design and fine tuning of several aspects of our upcoming 
photometric survey, such as the definition of the best observational strategy, the optmization of the target list, 
and the identification of improvements to be carried out on the pipeline for the photometric data reduction and 
time-series periodicity analysis. Our main findings can be summarized as follows: 

$\bullet$ $\textit{Photometric precision}$. We achieve a typical nightly RMS photometric
precision of $\sim5$ mmag, with little or no dependence on the instrumentation adopted or on
the details of the methodology (different comparison stars selection criteria, use of different 
detrending algorithms) utilized to perform differential photometry on the targets. 
We also carried out an analysis of the impact of correlated (red) noise on time-scales of $\backsim$ 30
min, which showed that it is typically a factor $\backsim$ 1.3 greater
than pure white noise, with a weak dependence on the method used to
perform differential photometry. This result reveals that our data are only mildly affected by short-term correlated systematics. 
The estimated photometric precision degrades to $\sim9$ mmag when the ensemble 
light curves are determined over the typical $\sim2$ months duration of the observations for each target. 
Such degradation is understood in terms of a combination of unmodeled medium-term systematics in our data 
and intrinsic variability of our target stars.

$\bullet$ $\textit{Stellar variability analysis}$. 
We searched for periodic transit-like events in the photometric dataset 
for each target using the BLS algorithm. 
No such signal was recovered for any target. This is an expected result given the sample size, 
thus meaningful constraints/upper limits on the planet fraction as a function of radius and orbital 
separation cannot be provided. The light curves of our program stars were inspected for evidence of 
periodic signals of approximately sinusoidal shape, which could be interpreted 
as due to the presence of rotating spots on the stellar photosphere. 
For two stars in our sample, LHS 3445 and GJ 1167A, we found clear evidence of a
periodicity in the light curve ascribable to such effect. 
We determined photometric rotation periods of $\sim0.47$ days and $\sim0.22$ days for 
LHS 3445 and GJ 1167A, respectively; these estimates were confirmed by the large
projected rotational velocities ($v\sin i \sim 25$ km/s and $v\sin i\sim 33$ km/s, 
respectively) inferred for both stars based on the analysis of archival high-resolution 
Keck/HIRES spectra. The estimated inclinations of the
stellar rotation axes for LHS 3445 and GJ 1167A agree with those derived using a simple
spot model, which successfully reproduces the observed sinusoidal photometric variations in
both cases (the dispersion of the post-fit residuals is on the order of the sample photometric precision). 
Finally, we detected short-term, low-amplitude flaring events in the differential 
photometric measurements of LHS 3445 and LHS 2686 (the latter not known to be a flare star). 
LHS 3445 was observed flaring three times, and two flares were recorded almost
consecutively during the same night, with an approximately equal decay
time of $\backsim$ 4.5 min, possibly a case of homologous flares.

$\bullet$ $\textit{Sensitivity to small-radius transiting planets}$. We 
carried out large-scale simulations of transit signals (of periods in the range 
$0.5-5$ days and depths in the range 0.5\%-2\% in flux units) injected 
in the actual (nightly reduced) photometric data for our sample. A total of 
400,000 light curves were analysed for each target using a real-life transit 
events search algorithm (BLS). The study of the BLS transit recovery rates and overall 
performance for a sub-sample of stars with good, fair, and poor phase coverage 
highlighted the capability of BLS to identify the correct period (when multiple transits 
were observed) even for signals 
with depth close to the typical photometric precision of the data ($\sim5$ mmag), 
albeit with low statistical confidence, as well as some of its performance limitations 
which are driven by the specific choice of its most relevant setup parameters. 
We expressed our main findings in terms of two easy-to-use comparison metrics, 
i.e. transit detection probabilities and phase coverage. We found a quasi-linear 
relationship between the two quantities. Based on the BLS algorithm, there appears to be 
a limit of $\approx 90\%$ in the probability of detecting a transit even when the 
phase coverage approaches $100\%$. Around stars in our sample with
good phase coverage ($> 50\%$), we would have had $> 80\%$ chances of detecting companions with
$P < 1$ day and transit depths $>0.5\%$ in flux units. Correspondingly, 
around these stars we would have been sensitive to companions with radii as small as $\sim1.0-2.2$ $R_\oplus$. 

The main findings reported here 
provide useful information for the purpose of the design and implementation of the operations of a ground-based M dwarf transit survey, with the 
aim of maximizing the chances for small-radius planet detection and improving our understanding of several astrophysically 
interesting properties of M dwarfs, particularly when investigated by means of statistical analyses of large stellar samples.
Together with other similar efforts carried out by other groups, such as the pioneering MEarth program, the photometric database 
populated by our survey data will be of great help, for example, a) to improve the characterization 
of nearby M dwarf stars, when combined with Gaia's exquisitely accurate astrometry (e.g., Sozzetti 2011), 
and b) to optimize the target selection criteria for red dwarfs which might be included in next-generation space-based transit 
survey programs, such as TESS (Ricker et al. 2010) and PLATO (Rauer \& Catala 2011) or which might be selected  
for spectroscopic characterization of planetary atmospheres of transiting planets found orbiting cool, nearby stars 
with future space-borne infrared observatories such as EChO (Tinetti et al. 2011) and FINESSE (Swain 2010).

A forthcoming paper will present in detail all relevant aspects of the upcoming survey, including 
overall systems description, operations control software, target selection criteria, robust reduction 
pipeline and archiving. 

\section*{Acknowledgments}

We warmly thank the anonymous referee for providing very useful comments and insightful suggestions 
which greatly helped to materially improve the paper. 
This research has made use of the SIMBAD database, operated at CDS, Strasbourg, France, 
and of the Keck Observatory Archive (KOA), which is operated by the W. M. Keck Observatory and the NASA Exoplanet Science Institute (NExScI), 
under contract with the National Aeronautics and Space Administration. 
MD, PC, AB, and GT are supported by grants of the European Union, the Autonomous Region 
of the Aosta Valley and the Italian Department for Work, Health and Pensions. 
The OAVdA is supported by the Regional Government of Valle d'Aosta, the Town Municipality of Nus and the Monte Emilius Community.
We thank Davide Cenadelli (OAVdA) for his contribution in the analysis of the Keck/HIRES archival spectra.

\label{lastpage}


\begin{thebibliography}{999}

\bibitem[Abell et al., 2009]{abell09}
Abell, P., Allison, J., Anderson, S. F., Andrew, J. R., Angel, J. R. P., Armus, L., Arnett, D., Asztalos, S. J., et al. 2009, LSST Science Book (arXiv:0912.0201)
\bibitem[Adams et al., 2008]{adams08}
Adams, E. R., Seager, S., Elkins-Tanton, L. 2008, \apj, 673, 1160
\bibitem[Aerts et al., 2010]{aerts10}  
Aerts, C., Christensen-Dalsgaard, J., and Kurtz, D. W., 2010, Asteroseismology, Springer Ed.
\bibitem[Ambartsumyan et al., 1970]{ambar70}
Ambartsumyan, V. A., Mirzoyan, L. V., Parsamyan, E. S., Chavushyan, O. S., Erastova, L. K. 1970, Astrofizika, 6, 7
\bibitem[Apps et al., 2010]{apps10}
Apps, K., Clubb, K. I., Fischer, D. A., Gaidos, E., Howard, A., Johnson, J.
A., Marcy, G. W., Isaacson, H., et al. 2010, \pasp, 122, 156
\bibitem[Audard et al., 2000]{audard00}
Audard, M., G\"udel, M., Drake, J. J., Kashyap, V. L. 2000, \apj, 541, 396
\bibitem[Bailer-Jones \& Lamm, 2003]{bailer03}
Bailer-Jones, C. A. L., Lamm, M. 2003, \mnras, 339, 477
\bibitem[Barnes et al., 2011]{barnes11}
Barnes, J.R., Jeffers, S.V., Jones, H.R.A. 2011, \mnras, 412, 1599
\bibitem[Bayless et al., 2006]{bayless06}
Bayless, A. J., \& Orosz, J. A. 2006, \apj, 651, 1155
\bibitem[Bean et al., 2006]{bean06}
Bean, J. L., Sneden, C., Hauschildt, P. H., Johns-Krull, C. M., Benedict, G. F. 2006, \apj, 652, 1604
\bibitem[Bean et al., 2010]{bean10}
Bean, J. L., Miller-Ricci Kempton, E., Homeier, D. 2010, \nat, 468, 669
\bibitem[Bean et al., 2011]{bean11}
Bean, J. L., D\'esert, J.-M., Kabath, P., Stalder, B., Seager, S., Miller-Ricci Kempton, E., Berta, Z. K., Homeier, D., et al. 2011, \apj,\, 743, 92
\bibitem[Beatty et al., 2007]{beatty07}
Beatty, T. G., et al. 2007, \apj, 663, 573
\bibitem[Bochanski et al., 2005]{bochanski05}
Bochanski, J. J., Hawley, S. L., Reid, I. N., Covey, K. R., West, A. A., Tinney, C. G., Gizis, J. E. 2005, \aj, 130, 1871
\bibitem[Boisse et al., 2011]{boisse11}
Boisse, I., Bouchy, F., H\'ebrard, G., Bonfils, X., Santos, N., Vauclair, S. 2011, \aap, 528, A4
\bibitem[Bonfils et al., 2005]{bonfils05}
Bonfils, X., Delfosse, X., Udry, S., Santos, N. C., Forveille, T., S\'egransan, D. 2005, \aap, 442, 635
\bibitem[Bonfils et al., 2007]{bonfils07}
Bonfils, X., Mayor, M., Delfosse, X., Forveille, T., Gillon, M., Perrier, C., Udry, S., Bouchy, F., et al. 2007, \aap, 474, 293
\bibitem[Browning et al., 2010]{browning10}
Browning, M.~K., Basri, G., Marcy, G. W., West, A. A., Zhang, J.  2010, \aj, 139, 504 
\bibitem[Burke et al., 2006]{burke06}
Burke, C. J., Gaudi, B. S., DePoy, D. L., Pogge, R. W. 2006, \aj, 132, 210
\bibitem[Casagrande et al., 2008]{casagrande08}
Casagrande, L., Flynn, C., Bessell, M. 2008, \mnras, 389, 585
\bibitem[Charbonneau et al., 2009]{charbonneau09}
Charbonneau, D., Berta, Z. K., Irwin, J., Burke, C. J., Nutzman, P., Buchhave, L. A., Lovis, C., Bonfils, X., et al. 2009, \nat, 462, 891
\bibitem[Ciardi et al., 2011]{ciardi11}
Ciardi, D. R., von Braun, K., Bryden, G., van Eyken, J., Howell, S. B., Kane, S. R., Plavchan, P., Ram\'irez, S. V., et al. 2011, \aj, 141, 108 
\bibitem[Claret, 2000]{claret00}
Claret, A. 2000, \aap, 363, 1081
\bibitem[Croll et al., 2011]{croll11}
Croll, B., Albert, L., Jayawardhana, R., Miller-Ricci Kempton, E., Fortney, J. J., Murray, N., Neilson, H. 2011, \apj, 736, 78
\bibitem[Crossfield et al., 2011]{crossfield11}
Crossfield, I. J. M., Barman, T., Hansen, B. M. S. 2011, \apj, 736, 132
\bibitem[Damasso et al., 2010]{Dam10}
Damasso, M., Giacobbe, P., Calcidese, P., Sozzetti, A., Lattanzi, M. G., Bernagozzi, A., Bertolini, E., 
Smart, R. L. 2010, \pasp, 122, 1077
\bibitem[Deeming, 1975]{deeming}
Deeming, T. J. 1975, \apss, 36, 137 
\bibitem[Delfosse et al., 2000]{delfosse00}
Delfosse, X., Forveille, T., S\'egransan, D., Beuzit, J.-L., Udry, S., Perrier, C., Mayor, M. 2000, \aap, 364, 217
\bibitem[D\'esert et al., 2011]{desert11}
D\'esert, J.-M., Bean, J., Miller-Ricci Kempton, E., Berta, Z. K., Charbonneau, D., Irwin, J., Fortney, J., Burke, C. J., et al. 2011, \apj, 731, L40
\bibitem[Doyle et al., 1990]{doyle90}
Doyle, J. G., Butler, C. J., van den Oord, G. H. J., Kiang, T. 1990, \aap, 232, 93
\bibitem[Droege et al., 2006]{droege06}
Droege, T. F., Richmond, M. W., Sallman, M. 2006, PASP, 118, 1666
\bibitem[Dumusque et al., 2011]{dumusque11}
Dumusque, X., Santos, N. C., Udry, S., Lovis, C., Bonfils, X. 2011, \aap, 527, A82
\bibitem[Eggenberger \& Udry, 2010]{eggenberger10}
Eggenberger, A., Udry, S. 2010, EAS Pub. Ser., 41, 27
\bibitem[Endl et al., 2006]{endl06}
Endl, M., Cochran, W. D., K\"urster, M., Paulson, D. B., Wittenmyer, R. A., MacQueen, P. J., Tull, R. G. 2006, \apj, 649, 436
\bibitem[Ferraz-Mello et al., 2010]{ferrazmello10}
Ferraz-Mello, S., Tadeu Dos Santos, M., Beaug\'e, C., Michtchenko, T. A., Rodr{\'\i}guez, A. 2011, \aap, 531, A161
\bibitem[Fischer \& Valenti, 2005]{fischer05}
Fischer, D. A., Valenti, J. 2005, \apj, 622, 1102
\bibitem[Forveille et al., 2011]{forveille11}
Forveille, T., Bonfils, X., Lo Curto, G., Delfosse, X., Udry, S., Bouchy, F., Lovis, C., Mayor, M., et al. 2011, \aap, 526, A141
\bibitem[Gershberg et al., 1999]{gershberg99}
Gershberg, R.E., Katsova, M.M., Lovkaya, M.N., Terebizh, A.V., Shakhovskaya, N.I. 1999, \aaps, 139, 555
\bibitem[Gizis et al., 2002]{gizis02}
Gizis, J. E., Reid, I. N., Hawley, S. L. 2002, \aj, 123, 3356
\bibitem[Gregory, 2011]{gregory11}
Gregory, P. C. 2011, \mnras, 415, 2523
\bibitem[G\"udel et al., 2003]{gudel03}
G\"udel, M., Audard, M., Kashyap, V. L., Drake, J. J., Guinan, E. F. 2003, \apj, 582, 423
\bibitem[Gustafsson, 1989]{gustaf89}
Gustafsson, B. 1989, \araa, 27, 701
\bibitem[Guinan et al., 2009]{guinan10}
Guinan, E. F., Engle, S. G., Mizusawa, T., McCook, G. P., Wolfe, A., Coughlin, J. 2009, Highlights of Astronomy, 15, 703
\bibitem[Gray, 2008]{gray08}
Gray, D. F. 2008, The Observation and Analysis of Stellar Photospheres, by D. F. Gray. Cambridge: Cambridge University Press, 2008  
\bibitem[Haghighipour et al., 2010]{haghighipour10}
Haghighipour, N., Vogt, S. S., Butler, R. P., Rivera, E. J., Laughlin, G., Meschiari, S., Henry, G. W. 2010, \apj, 715, 271
\bibitem[Hartman et al., 2009]{hartman09}
Hartman, J. D., Gaudi, B. S., Pinsonneault, M. H., Stanek, K. Z., Holman, M. J., McLeod, B. A., Meibom, S., Barranco, J. A., et al. 2009, \apj, 691, 342
\bibitem[Hartman et al., 2011]{hartman11}
Hartman, J.D., Bakos, G.\'A., Noyes, R.W., Sipocz, B., Kov\'acs, G., Mazeh, T., Shporer, A., P\'al, A. 2011, \aj, 141, A166
\bibitem[Hatzes et al., 2010]{hatzes10}
Hatzes, A. P., Dvorak, R., Wuchterl, G., Guterman, P., Hartmann, M., Fridlund, M., Gandolfi, D., Guenther, E., et al. 2010, \aap, 520, A93
\bibitem[Henry et al., 2006]{henry06}
Henry, T. J., Jao, W., Subasavage, J. P., Beaulieu, T. D., Ianna, P. A., Costa, E., Méndez, R. A. 2006, AJ, 132, 2360
\bibitem[Howard et al., 2011]{howard11}
Howard, A.W., Marcy, G.W., Bryson, S.T., Jenkins, J.M., Rowe, J.F., Batalha, N.M., Borucki, W.J., Koch, D.G., et al. 2011, \apj\, submitted (arXiv:1103.2541)
\bibitem[Ida \& Lin, 2004]{ida04}
Ida, S., Lin, D. N. C. 2004, \apj, 616, 567
\bibitem[Ida \& Lin, 2005]{ida05}
Ida, S., Lin, D. N. C. 2005, \apj, 626, 1045
\bibitem[Irwin et al., 2007]{irwin07}
Irwin, J., Irwin, M., Aigrain, S., Hodgkin, S., Hebb, L., Moraux, E. 2007, \mnras, 375, 1449
\bibitem[Irwin et al., 2011]{irwin11}
Irwin, J., Berta, Z. K., Burke, C. J., Charbonneau, D., Nutzman, P., West, A. A., Falco, E. E. 2011, \apj, 727, 56
\bibitem[Jenkins et al., 2009]{jenkins09}
Jenkins, J. S., Ramsey, L. W., Jones, H. R. A., Pavlenko, Y., Gallardo, J., Barnes, J. R., Pinfield, D. J. 2009, \apj, 704, 975
\bibitem[Johnson et al., 2007]{johnson07}
Johnson, J. A., Fischer, D. A., Marcy, G. W., Wright, J. T., Driscoll, P., Butler, R. P., Hekker, S., Reffert, S., et al. 2007, \apj, 665, 785
\bibitem[Johnson \& Apps, 2009]{johnson09}
Johnson, J. A., Apps, K. 2009, \apj, 699, 933
\bibitem[Johnson et al., 2010]{johnson10}
Johnson, J. A., Aller, K. M., Howard, A. W., Crepp, J. R. 2010, \pasp, 122, 905
\bibitem[Johnson et al., 2011a]{johnson11a}
Johnson, J. A., Clanton, C., Howard, A. W., Bowler, B. P., Henry, G. W., Marcy, G. W., Crepp, J. R., Endl, M., et al. 2011a, \apjs, 197, 26
\bibitem[Johnson et al., 2012]{johnson12} 
Johnson, J. A., Gazak, J. Z., Apps, K., Muirhead, P. S., Crepp, J. R., Crossfield, I. J. M., Boyajian, T., von Braun, K., et al. 2012, \aj, 143, 111
\bibitem[Lenz \& Breger, 2005]{lenz05}
Lenz, P., \& Breger, M. 2005, Communications in Asteroseismology, 146, 53
\bibitem[Kaiser, 2004]{kaiser04}
Kaiser, N. 2004, in Ground-based Telescopes. Proc. SPIE, 5489, 11
\bibitem[Kasting et al., 1993]{kasting93}
Kasting, J. F., Whitmire, D. P., \& Reynolds, R. T. 1993, \icarus, 101, 108
\bibitem[Kennedy \& Kenyon, 2008]{kennedy08}
Kennedy, G. M., Kenyon, S. J. 2008, \apj, 682, 1264
\bibitem[Kharchenko, 2001]{khar01}
Kharchenko, N.V., 2001, Kinematika Fiz. Nebesn. Tel., 17, 409
\bibitem[Kov\'acs et al., 2002]{kovacs02}
Kov\'acs, G., Zucker, S., \& Mazeh, T. 2002, \aap, 391, 369-377
\bibitem[Kov\'acs et al., 2005]{kovacs05}
Kov\'acs, G., Bakos, G., Noyes, R. W. 2005, \mnras, 356, 557
\bibitem[Kowalski et al., 2009]{kowalski09}
Kowalski, A. F., Hawley, S. L., Hilton, E. J., Becker, A. C., West, A. A., Bochanski, J. J., Sesar, B. 2009, \aj, 138, 633
\bibitem[Krautter, 1996]{krautter96}
Krautter, J. 1996, in Light Curves of Variable Stars: a Pictorial Atlas, ed. C. Jaschek \& C. Sterken (Cambridge: Cambridge Univ. Press), 53
\bibitem[Lacy et al., 1976]{lacy76}
Lacy, C. H., Moffett, T. J., \& Evans, D. S. 1976, \apjs, 30, 85
\bibitem[Lagrange et al., 2010]{lagrange10}
Lagrange, A.-M., Desort, M., Meunier, N. 2010, \aap, 512, A38
\bibitem[Lammer et al., 2007]{lammer07}
Lammer, H., Lichtenegger, H. I. M., Kulikov, Y. N., Grießmeier, J.-M., Terada, N., Erkaev, N. V., Biernat, H. K., Khodachenko, M. L., et al. 2007, Astrobiology, 7, 185
\bibitem[Laughlin et al., 2004]{laughlin04}
Laughlin, G., Bodenheimer, P., Adams, F. C. 2004, \apj, 612, L73
\bibitem[Law et al., 2009]{law09}
Law, N. M., Kulkarni, S. R., Dekany, R. G., Ofek, E. O., Quimby, R. M., Nugent, P. E., Surace, J., Grillmair, C. C., et al. 2009, \pasp, 121, 1395
\bibitem[L\'epine \& Shara, 2005]{lep05} 
L\'epine, S. \& Shara, P. 2005, \aj, 129, 1438
\bibitem[L\'epine, 2005]{lepine05} 
L\'epine, S. 2005, \aj, 130, 1680
\bibitem[Levenberg, 1944]{levenberg44}
Levenberg, K. 1944, Quart. Appl. Math., 2, 164
\bibitem[L\'opez-Morales, 2007]{lopezm07}
L\'opez-Morales, M. 2007, \apj, Vol. 660, Issue 1, pp. 732-739
\bibitem[L\'opez-Santiago et al., 2010]{lopez10}
L\'opez-Santiago, J., Montes, D., G\'alvez-Ortiz, M. C., Crespo-Chac\'on, I., Mart\'inez-Arn\'aiz, R. M., Fern\'andez-Figueroa, M. J., de Castro, E., Cornide, M. 2010, \aap, 514, A97
\bibitem[Mandel \& Agol, 2002]{mandel02}
Mandel, K. \& Agol, E. 2002, \apj, 580, L171
\bibitem[Makarov et al., 2009]{makarov09}
Makarov, V.V., Beichman, C.A., Catanzarite, J.H., Fischer, D.A., Lebreton, J., Malbet, F., Shao, M., 2009, \apj, 707, L73
\bibitem[Markwardt, 2009]{markwardt09}
Markwardt, C. B. 2009, arXiv:0902.2850
\bibitem[Marquardt, 1963]{marquardt63}
Marquardt, D. 1963, SIAM J. Appl. Math., 11, 431
\bibitem[Martres et al., 1984]{martres84}
Martres, M.-J., Mein, N., Mouradian, Z., Rayrole, J., Schmieder, B., Simon, G., Soru-Escaut, I., Woodgate, B. E. 1984, AdSR, 4, 5
\bibitem[Meibom et al., 2009]{meibom09}
Meibom, S., Mathieu, R. D., Stassun, K. G. 2009, \apj, 695, 679
\bibitem[Miller-Ricci \& Fortney, 2010]{miller10}
Miller-Ricci, E., \& Fortney, J.J. 2010, \apj, 716, L74
\bibitem[Miller-Ricci et al., 2011]{miller11}
Miller-Ricci Kempton, E., Zahnle, K., Fortney, J. J. 2011, \apj, 745, 3
\bibitem[Mirzoyan et al., 1989]{mirzoyan89}
Mirzoyan, L. V., Ambaryan, V. V., Garibdzhanyan, A. T., Mirzoyan, A. L. 1989, Astrophysics, 31, 567
\bibitem[Moffett, 1974]{moffett74}
Moffett, T. J. 1974, \apjs, 29, 1
\bibitem[Mordasini et al., 2009]{mordasini09}
Mordasini, C., Alibert, Y., Benz, W., Naef, D. 2009, \aap, 501, 1161
\bibitem[Nettelmann et al., 2011]{nettel11}
Nettelmann, N., Fortney, J. J., Kramm, U., Redmer, R. 2011, \apj, 733, 2
\bibitem[Nutzman \& Charbonneau, 2008]{nutzman08}
Nutzman, P., \& Charbonneau, D. 2008, \pasp, 120, 317
\bibitem[Pepe et al., 2011]{pepe11}
Pepe, F., Mayor, M., Lovis, C., Benz, W., Bouchy, F., Dumusque, X., Queloz, D., Santos, N. C., et al. 2011, Proc. IAU Symp. 276, 13
\bibitem[Phan-Bao et al., 2009]{phanbao09}
Phan-Bao, N., Lim, J., Donati, J.-F., Johns-Krull, C. M., Mart{\'{\i}}n, E. L. 2009, \apj, 704, 1721 
\bibitem[Pizzolato et al., 2003]{pizzolato03}
Pizzolato, N., Maggio, A., Micela, G., Sciortino, S., Ventura, P. 2003, \aap, 397, 147
\bibitem[Pont et al., 2006]{pont06}
Pont, F., Zucker, S., Queloz, D. 2006, \mnras, 373, 231
\bibitem[Pont et al., 2011]{pont11}
Pont, F., Aigrain, S., Zucker, S. 2011, \mnras, 411, 1953
\bibitem[Queloz et al., 2001]{queloz01}
Queloz, D., Henry, G. W., Sivan, J. P., Baliunas, S. L., Beuzit, J. L., Donahue, R. A., Mayor, M., Naef, D., et al. 2001, \aap, 379, 279
\bibitem[Queloz et al., 2009]{queloz09}
Queloz, D., Bouchy, F., Moutou, C., Hatzes, A., H\'ebrard, G., Alonso, R., Auvergne, M., Baglin, A., et al. 2009, \aap, 506, 303
\bibitem[Ranns et al., 2000]{ranns00}
Ranns, N. D. R., Harra, L. K., Matthews, S. A., Culhane, J. L. 2000, \aap, 360, 1163
\bibitem[Rauer \& Catala, 2011]{rauer11}
Rauer, H., \& Catala, C. 2011, Proc. IAU Symp. 276, 354
\bibitem[Reid \& Mahoney, 2000]{reid2000}
Reid, I. N., Mahoney, S. 2000, \mnras, 316, 827
\bibitem[Reid \& Cruz, 2002]{reid02}
Reid I.N., Cruz K.L. 2002,  Astron. J. 123, 2006
\bibitem[Reiners, 2007]{reiners07}
Reiners, A. 2007, \aap, 467, 259
\bibitem[Reiners \& Basri, 2010]{reiners10}
Reiners, A., Basri, G. 2010, \aap, 710, 924
\bibitem[Reiners et al., 2010]{reinersetal10}
Reiners, A., Bean, J. L., Huber, K. F., Dreizler, S., Seifahrt, A., Czesla, S. 2010, 710, 432
\bibitem[Ribas, 2006]{ribas06}
Ribas, I. 2006, \apss, 304, 89
\bibitem[Ricker et al., 2010]{ricker10}
Ricker, G. R., Latham, D. W., Vanderspek, R. K., Ennico, K. A., Bakos, G., Brown, T. M., Burgasser, A. J., Charbonneau, D., et al. 2010, BAAS, 42, 459
\bibitem[Rojas-Ayala et al., 2010]{rojas10}
Rojas-Ayala, B., Covey, K. R., Muirhead, P. S., Lloyd, J. P. 2010, \apj, 720, L113
\bibitem[Rogers \& Seager, 2010]{rogers10}
Rogers, L. A., Seager, S. 2010, \apj, 716, 1208
\bibitem[Santos et al., 2004]{santos04}
Santos, N. C., Israelian, G., Mayor, M. 2004, \aap, 415, 1153
\bibitem[Scalo et al., 2007]{scalo07}
Scalo, J., Kaltenegger, L., Segura, A.G., Fridlund, M., Ribas, I., Kulikov, Yu.N., Grenfell, J.L., Rauer, H., et al. 2007, Astrobiology, 7, 85
\bibitem[Schlaufman \& Laughlin, 2010]{schlauf10}
Schlaufman, K. C., Laughlin, G. 2010, \aap, 519, A105
\bibitem[Seager \& Mallen-Ornelas, 2003]{seager03}
Seager, S. \& Mallen-Ornelas, G. 2003, \apj, 585, 1038
\bibitem[Seager \& Deming, 2010]{seager10}
Seager, S., Deming, D. 2010, \araa, 48, 631
\bibitem[Shkolnik et al., 2009]{shkolnik09}
Shkolnik, E., Liu, M. C., Reid, I. N. 2009, \apj, 699, 649
\bibitem[Smart et al., 2010]{smart10}
Smart, R. L., Ioannidis, G., Jones, H. R. A., Bucciarelli, B., \& Lattanzi, M. G. 2010, \aap, 514, A84
\bibitem[Sozzetti et al., 2009]{sozzetti09}
Sozzetti, A., Torres, G., Latham, D. W., Stefanik, R. P., Korzennik, S. G., Boss, A. P., Carney, B. W., Laird, J. B. 2009, \apj, 697, 544
\bibitem[Sozzetti, 2011]{sozzetti11}
Sozzetti, A. 2011, EAS Pub. Ser., 45, 273
\bibitem[Stellingwerf, 1978]{stellingwerf78}
Stellingwerf, R. F. 1978, \apj, 224, 953 
\bibitem[Strassmeier et al., 2000]{strass00}
Strassmeier, K., Washuettl, A., Granzer, Th., Scheck, M., Weber, M. 2000, \aaps, 142, 275
\bibitem[Swain, 2010]{swain10}
Swain, M. R. 2010, BAAS, 42, 1064
\bibitem[Tamuz et al., 2005]{tamuz05}
Tamuz, O., Mazeh, T., Zucker, S. 2005, \mnras, 256, 1466
\bibitem[Tarter et al., 2007]{tarter07}
Tarter, J.C., Backus, P.R., Mancinelli, R.L., Aurnou, J.M.,
Backman, D.E., Basri, G.S., Boss, A.P., Clarke, A., et al. 2007, Astrobiology, 7, 30
\bibitem[Tinetti et al., 2011]{tinetti11}
Tinetti, G., Beaulieu, J.P., Henning, T., Meyer, M., Micela, G., Ribas, I., Stam, D., Swain, M., et al. 2011, Exp. Ast., accepted (arXiv:1112.2728) 
\bibitem[Torres et al., 2010]{torres10}
Torres, G., Andersen, J., Gim\'enez, A. 2010, A\&ARv, 18, 67
\bibitem[Tuomi, 2011]{tuomi11}
Tuomi, M. 2011, \aap, 528, L5
\bibitem[Van Leeuwen, 2007]{vanleewen07}
Van Leeuwen, F. 2007, \aap, 474, 653
\bibitem[Vogt et al., 2010]{vogt10}
Vogt, S. S., Butler, R. P., Rivera, E. J., Haghighipour, N., Henry, G. W., Williamson, M. H. 2010, \apj, 723, 954
\bibitem[Walkowicz \& Hawley, 2009]{walkowicz09}
Walkowicz, L.~M., \& Hawley, S.L. 2009, \aj, 137, 3297
\bibitem[West \& Basri, 2011]{west09}
West, A. A., Basri, G. 2009, \apj, 693, 1283
\bibitem[West et al., 2011]{west11}
West, A. A., Bochanski, J. J., Bowler, B. P., Dotter, A., Johnson, J. A., L\'epine, S., Rojas-Ayala, B., Schweitzeret A. 2011, ASP. Conf. Ser., 448, 531
\bibitem[Woolf \& Wallerstein, 2006]{woolf06}
Woolf, V. M., Wallerstein, G. 2006, \pasp, 118, 218
\bibitem[Woolf et al., 2009]{woolf09}
Woolf, V. M., L\'epine, S., Wallerstein, G. 2009, \pasp, 121, 117
\bibitem[Wright et al., 2004]{wright04}
Wright, J. T., Marcy, G. W., Butler, R. P., Vogt, S. S. 2004, \apjs, 152, 261
\bibitem[Zechmeister et al., 2009]{zech09}
Zechmeister, M., K\"urster, M., Endl, M. 2009, \aap, 505, 859
\bibitem[Xucker \& Mazeh, 1994]{zucker94}
Zucker, S., \& Mazeh, T. 1994, \apj, 420, 806

\end{thebibliography}
\end{document}